\documentclass[12pt]{article}

\usepackage{graphicx}
\usepackage{graphics}
\usepackage{amssymb}
\usepackage{amsmath}

\usepackage{epsf,amsfonts,hyperref}

\renewcommand{\appendix}[1]{
    \addtocounter{section}{1}
    \setcounter{equation}{0}
    \renewcommand{\thesection}{\Alph{section}}
    \section*{Appendix \thesection\protect\indent #1}
    \addcontentsline{toc}{section}{Appendix \thesection\ \ \ #1}
}
\newcommand\encadremath[1]{\vbox{\hrule\hbox{\vrule\kern8pt 
\vbox{\kern8pt \hbox{$\displaystyle #1$}\kern8pt} 
\kern8pt\vrule}\hrule}}
\def\enca#1{\vbox{\hrule\hbox{
\vrule\kern8pt\vbox{\kern8pt \hbox{$\displaystyle #1$}
\kern8pt} \kern8pt\vrule}\hrule}}

\newcommand\figureframex[3]{
\begin{figure}[bth]
\hrule\hbox{\vrule\kern8pt 
\vbox{\kern8pt \vbox{
\begin{center}
{\mbox{\epsfxsize=#1.truecm\epsfbox{#2}}}
\end{center}
\caption{#3}
}\kern8pt} 
\kern8pt\vrule}\hrule
\end{figure}
}
\newcommand\figureframey[3]{
\begin{figure}[bth]
\hrule\hbox{\vrule\kern8pt 
\vbox{\kern8pt \vbox{
\begin{center}
{\mbox{\epsfysize=#1.truecm\epsfbox{#2}}}
\end{center}
\caption{#3}
}\kern8pt} 
\kern8pt\vrule}\hrule
\end{figure}
}

\makeatletter
\@addtoreset{equation}{section}
\makeatother
\newtheorem{theorem}{Theorem}[section]

\newtheorem{remark}{Remark}[section]
\newtheorem{proposition}{Proposition}[section]
\newtheorem{lemma}{Lemma}[section]
\newtheorem{corollary}{Corollary}[section]
\newtheorem{definition}{Definition}[section]
\def\br{\begin{remark}\rm\small}
\def\er{\end{remark}}
\def\bt{\begin{theorem}}
\def\et{\end{theorem}}
\def\bd{\begin{definition}}
\def\ed{\end{definition}}
\def\bp{\begin{proposition}}
\def\ep{\end{proposition}}
\def\bl{\begin{lemma}}
\def\el{\end{lemma}}
\def\bc{\begin{corollary}}
\def\ec{\end{corollary}}
\def\beaq{\begin{eqnarray}}
\def\eeaq{\end{eqnarray}}
\newcommand{\proof}[1]{{\noindent \bf proof:}\par
{#1} $\square$}

\newcommand{\eq}[1]{Eq.~(\ref{#1})}

\newcommand{\beq}{\begin{equation}}
\newcommand{\eeq}{\end{equation}}
\newcommand{\bea}{\begin{eqnarray}}
\newcommand{\eea}{\end{eqnarray}}

\renewcommand{\and}{{\qquad {\rm and} \qquad}}

\newcommand{\virg}{{\qquad , \qquad}}

\newcommand{\Res}{\mathop{\,\rm Res\,}}

\newcommand{\td}[1]{{\tilde{#1}}}

\newcommand{\om}{\omega}

\newcommand{\ee}[1]{{{\rm e}^{#1}}}

\renewcommand{\d}{{{\partial}}}

\newcommand{\Pint}{{\int\kern -1.em -\kern-.25em}}

\renewcommand{\Im}{{\mathrm{Im}}}

\renewcommand{\L}{\Lambda}

\newcommand{\bcycle}{{\cal B}}

\newcommand\spcurve{{\cal S}}
\newcommand\curve{{\cal C}}

\newcommand\Brond{{\stackrel{\circ}{B}}}

\newcommand\Wrond{{\stackrel{\circ}{W}}}

\newcommand\spcurverond{{\stackrel{\circ}{\spcurve}}}

\newcommand\Ber{{{\cal B}}}

\begin{document}
\sloppy


\pagestyle{empty}
\hfill CERN-068

\hfill IPHT-T11/045
\addtolength{\baselineskip}{0.20\baselineskip}
\begin{center}
\vspace{26pt}
{\large \bf {Intersection numbers of spectral curves}}
\newline
\vspace{26pt}

{\sl B.\ Eynard}\hspace*{0.05cm}\footnote{ E-mail: eynard@spht.saclay.cea.fr }\\
\vspace{6pt}
CERN,\\
Service de Physique Th\'{e}orique de Saclay,\\
F-91191 Gif-sur-Yvette Cedex, France.\\
\end{center}

\vspace{20pt}
\begin{center}
{\bf Abstract}

We compute the symplectic invariants of an arbitrary spectral curve with only 1 branchpoint in terms of integrals of characteristic classes in the moduli space of curves.
Our formula associates to any spectral curve, a characteristic class, which is determined by the laplace transform of the spectral curve. This is a hint to the key role of Laplace transform in mirror symmetry.
When the spectral curve is $y=\sqrt x$, the formula gives Kontsevich--Witten intersection numbers, when the spectral curve is chosen to be the Lambert function $\ee{x}=y\ee{-y}$, the formula gives the ELSV formula for Hurwitz numbers, and when one chooses the mirror of $\mathbb C^3$ with framing $f$, i.e. $\ee{-x}=\ee{-yf}\,(1-\ee{-y})$, the formula gives the topological vertex Mari\~no--Vafa formula, i.e. the generating function of Gromov-Witten invariants of $\mathbb C^3$.
In some sense this formula generalizes ELSV, Mari\~no--Vafa formula, and Mumford formula.

\end{center}
%





\vspace{26pt}
\pagestyle{plain}
\setcounter{page}{1}


\section{Introduction}

In \cite{EOFg} were introduced some "symplectic invariants" of a spectral curve (we call spectral curve a plane curve, i.e. a Riemann surface embedded into $\mathbb C^2$, often chosen as the locus of zeroes of an analytical function $E(x,y)=0$).
Those invariants play an important role in random matrix theory, and in many enumerative geometry problems. They were first introduced in relationship with random matrices and enumeration of discrete surfaces. Indeed, if the spectral curve $\spcurve$ is chosen as the spectral curve of a matrix model, then the $g^{\rm th}$ symplectic invariant $F_g(\spcurve)$ is the $g^{\rm th}$ term in the large size expansion of the matrix integral, and it is the generating function enumerating discrete surfaces of genus $g$ (in fact this is the property which initially motivated the definition of symplectic invariants \cite{Eynard2004}).

Later it was realized that symplectic invariants of the spectral curve $y^2=x$ are the generating function of intersection numbers of Chern classes in ${\cal M}_{g,n}$ (the moduli space of curves of genus $g$ with $n$ marked points), i.e. Witten-Kontsevich intersection numbers \cite{Eynard}.

Then it was realized that they can also encode Gromov-Witten invariants \cite{Mar1,BKMP}.
If $\mathfrak X$ is a toric Calabi--Yau 3--fold, and $\spcurve=\hat{\mathfrak X}$ is its mirror singular curve (the mirror \cite{HoriKentaro2000} of a toric CY3, is a CY3 whose singular locus is a plane curve, which we call the mirror curve $\hat{\mathfrak X}$), then it was conjectured by Mari\~no \cite{Mar1} and then more precisely by Bouchard--Mari\~no--Klemm--Pasquetti in \cite{BKMP}, that $F_g(\hat{\mathfrak X})$ is the generating function of Gromov-Witten invariants counting stable maps of genus $g$ into $\mathfrak X$ (BKMP conjecture \cite{BKMP}). That conjecture was proved in a few cases \cite{ChenLin2009, ZhouJian2009}.

\smallskip

So we see that for "good" choices of spectral curves, the symplectic invariants have a beautiful enumerative geometry interpretation, they count some "number" of surfaces, or some "intersection numbers"  in the moduli space of curves or maps.

However, for an arbitrary spectral curve, different from the "good ones" listed above, it was so far not really known what symplectic invariants were counting.

\bigskip

Here in this article, we relate the symplectic invariants of an arbitrary spectral curve, to some intersection numbers.
Our formula is reminiscent of the ELSV formula \cite{EKEDAHL1999, Ekedahl2001} (relating Hurwitz numbers to intersection numbers of one Hodge class), or Mari\~no--Vafa formula \cite{MV01} for the vertex case, and the Mumford formula \cite{Mumford1983} (which rewrites the Hodge class in terms of $\psi$ and $\kappa$ classes).

\smallskip
Our formula for only one branchpoint is given by theorem \ref{thspinv1bp} below, which we write here:

{\bf Theorem} \ref{thspinv1bp}
\beq
W_n^{(g)}(\spcurve_a;z_1,\dots,z_n) = 2^{d_{g,n}}\!\!\!\!\sum_{d_1+\dots+d_n\leq d_{g,n}}\, \prod_i d\xi_{d_i}(z_i)\,\,\left< \ee{{1\over 2}\sum_\delta l_{\delta*}\hat B(\psi,\psi')}\,\ee{\sum_k \td t_{k} \kappa_k}\prod_i \psi_i^{d_i}\right>_{{g,n}}
\eeq
where the notations will be defined below.
To stay in an introductory level, we just point out that the left hand side of that formula is defined from the geometry of a spectral curve, i.e. a complex plane curve (therefore a type B quantity), whereas the right hand side contains intersection numbers of homology classes in some moduli-space, i.e. a type A quantity.
This looks like a kind of mirror symmetry \cite{HoriKentaro2000}, where the "mirror map" relates the B--moduli of the spectral curve, to the type A moduli $\td t_k$, $d\xi_d$, $\hat B$ in the right hand side, by Laplace transform.

\medskip
Let us now define our notations.

\section{Symplectic invariants of spectral curves}

\subsection{Spectral curves}

Intuitively, a spectral curve $\spcurve$ is a plane curve, i.e. the locus of solutions of an analytical function $E(x,y)=0$ in $\mathbb C^2$. In particular, it defines a Riemann surface $\curve$, and two projections $x:\curve\to \mathbb C$ and $y:\curve\to \mathbb C$.
By definition $\{ (x,y)\,| \, E(x,y)=0\}=\{(x(z),y(z))\,|\,\,z\in\curve\}$.
For example for the spectral curve $y^2=x$, we have $\{(x,y)\,|\,y^2=x\}=\{(z^2,z)\,|\,z\in\mathbb C\}$.

For our purposes, it is more convenient to define the spectral curve directely from the parametrization
$(\curve,x,y)$ where $\curve$ is a complex curve (a Riemann surface, not necessarily compact), and $x$ and $y$ are two analytical functions $\curve\to \mathbb C$.

\medskip

\bd
A spectral curve $\spcurve$, is the data of:
\beq
\spcurve=(\curve,x,y,B)
\eeq

$\bullet$ $\curve$ is a complex curve (a Riemann surface, not necessarily compact), 

$\bullet$ $x$ and $y$ are two analytical functions $\curve\to \mathbb C$.

$\bullet$ $B(z,z')$ is a "Bergman kernel", i.e. a symmetric 2nd kind differential on $\curve\times\curve$, having a double pole at $z=z'$ and no other pole, and which behaves like
\beq
B(z,z') \mathop{\sim}_{z\to z'}\,\, {dz\,\otimes\,dz'\over (z-z')^2} + O(1)
\eeq
in any local parameter $z$.
(In general the Bergman kernel is not unique, one may add to it anything which is holomorphic (with no pole) in $\curve\times\curve$ and symmetric).

\ed

\medskip
$\bullet$  An important example of spectral curve is:
\beq
y^2=x
\eeq
which can be parametrized by two functions $x(z)=z^2$ and $y(z)=z$ for a complex variable $z\in\curve=\mathbb C$,
and with the Bergman kernel
\beq
B(z,z')={dz\otimes dz'\over (z-z')^2}.
\eeq
In other words
\beq
\spcurve=(\mathbb C,x(z)=z^2,y(z)=z,B).
\eeq
This spectral curve, often called "Airy curve"\footnote{It is often called Airy curve because its symplectic invariants can be written in terms of Airy function. For instance $\sum_g W_1^{(g)}(z)  = (Ai'(z^2)Bi'(z^2)-z^2 Ai(z^2)Bi(z^2))2zdz = -\sum_g (6g-3)!!/(g!\,3^g\,2^{5g-1})\, z^{2-6g}dz$.} plays an important role in Witten--Kontsevich theory, as we shall see below.

\medskip
$\bullet$  Another interesting example is:
\beq
\ee{x}=y\,\ee{-y},
\eeq
called the "Lambert curve", because $y=L(\ee{x})$ where $L$ is the Lambert function.
It can be parametrized by $x(z)=-z+\ln{z}$, $y(z)=z$, $z\in \curve = \mathbb C^*\setminus \mathbb R_-$,
and the Bergman kernel is again chosen to be
\beq
B(z,z')={dz\otimes dz'\over (z-z')^2}.
\eeq
Our spectral curve is thus
\beq
\spcurve=(\mathbb C^*\setminus \mathbb R_-,x(z)=-z+\ln z,y(z)=z,B).
\eeq
This spectral curve plays an important role in Hurwitz numbers counting (as was noticed by Bouchard and Mari\~no \cite{BM} and proved in \cite{Eynardb, Borot1}), and reproved below in section \ref{secHurwitz} as a consequence of theorem \ref{thspinv1bp}.

\medskip
$\bullet$  Another interesting example is:
\beq
\ee{-x}=\ee{-fy}\,(1-\ee{-y}),
\eeq
called the "topological vertex curve", because it is the mirror curve of the framed topological vertex ($f\in\mathbb Z$ is the framing), indeed writing $X=\ee{-x}$ and $Y=\ee{-y}$, it satisfies:
\beq
X=Y^f\,(1-Y).
\eeq
It can be parametrized by $x(z)=- f\,\ln z-\ln{(1-z)}$, $y(z)=-\ln z$, $z\in \curve = \mathbb C^*\setminus (-\infty,0]\cup[1,\infty)$,
and the Bergman kernel is again chosen to be
\beq
B(z,z')={dz\otimes dz'\over (z-z')^2}.
\eeq
Our spectral curve is thus
\beq
\spcurve=(\mathbb C^*\setminus(-\infty,0]\cup[1,\infty),x(z)=-f\,\ln z-\ln{(1-z)},y(z)=-\ln z,B),
\eeq
This spectral curve plays an important role in the Gromov-Witten theory of $\mathbb C^3$ as was noticed by \cite{BKMP} and proved in \cite{ChenLin2009, ZhouJian2009}, and reproved below in section \ref{secvertex} as a consequence of theorem \ref{thspinv1bp}.

\subsection{Branchpoints}

A branchpoint is a zero of $dx$, i.e. $dx(a)=0$.
Let us assume that $a$ is a regular branchpoint, i.e. it is a simple zero of $dx$, and we have $dy(a)\neq 0$.
This means that locally near $a$ the curve has a square-root behavior:
\beq
y(z)\sim y(a)+y'(a)\,\sqrt{x(z)-x(a)} + O(x(z)-x(a))
\eeq
or also, that $\zeta(z)=\sqrt{x(z)-x(a)}$ is a good local parameter near $a$.

\medskip

In the vicinity of $a$, the square root has two branches, and we denote $\bar z$ the point corresponding to the other sign of the square-root, i.e. the unique other point in the vicinity of $a$ such that $\zeta(\bar z)=-\zeta(z)$, or also
\beq
x(\bar z)=x(z).
\eeq
Notice that $\bar z$ is well defined only when $z$ lies in the vicinity of a branchpoint, it is (in general) not globally defined for any $z\in \curve$.

Near a branchpoint $a$, we can Taylor expand, and define the "Kontsevich times":
\bd
The times $t_{k}$ of a branchpoint $a$, are the coefficients of the Taylor series:
\beq
y(z) \sim \sum_{k=0}^\infty\, t_{k+2}\,\zeta^k
\virg \zeta=\sqrt{x(z)-x(a)}.
\eeq
\ed

\smallskip
$\bullet$ Example for the Airy spectral curve $\spcurve=(\mathbb C,x(z)=z^2,y(z)=z,B)$, we have $dx(z)=2z dz$, which vanishes at $z=a=0$, there is only one branchpoint.
We clearly have $\bar z=-z$, which in this case is globally defined.
In that case, we have the times:
\beq
t_{k} = \delta_{k,3}.
\eeq

\smallskip
$\bullet$ Example for the Lambert spectral curve $\spcurve=(\mathbb C^*\setminus \mathbb R_-,x(z)=-z+\ln z,y(z)=z,B)$, we have $dx(z)={1-z\over z}\, dz$, which vanishes at $z=a=1$, there is only one branchpoint.
Near $z=1$ we have $\bar z= 2-z + {2\over 3}\, (z-1)^2+\dots O(z-1)^3$, which in this case is not globally defined.
In that case the times $t_k$ are given by
\beq
y = 1+i\sqrt 2\,\zeta - {2\over 3}\,\zeta^2+{11\,i\,\sqrt 2\over 9}\,\zeta^3+\dots \virg \zeta=\sqrt{x+1}.
\eeq

\subsection{Symplectic invariants}

Symplectic invariants were introduced in \cite{EOFg}.
For a spectral curve $\spcurve=(\curve,x,y,B)$, we define:
\bd
The "symplectic invariant descendents" $W_n^{(g)}(\spcurve;z_1,\dots,z_n)$ with $n\geq 1$ and $g\geq 0$ are defined by:
\beq
W_1^{(0)}(z) = y(z)\,dx(z)\,,
\eeq
\beq
W_2^{(0)}(z,z') = B(z,z')\,,
\eeq
and for $2g-2+n>0$, by the "topological recursion" \cite{Eynard2004, Alexandrov2005}
\bea
W_{n+1}^{(g)}(z_0,\overbrace{z_1,\dots,z_n}^{J})
&=& \sum_{a={\rm branch\, points}}\Res_{z\to a} K(z_0,z)\,\Big[ W_{n+2}^{(g-1)}(z,\bar z,J) \cr
&& + \sum_{h=0}^g\sum_{I\subset J}'\,W_{1+\#I}^{(h)}(z,I)\,W_{1+n-\#I}^{(g-h)}(\bar z,J\setminus I) \Big]
\eea
where $\sum'$ means that we exclude from the sum all terms which contain a factor $W_1^{(0)}$, and the recursion kernel $K$ is:
\beq
K(z_0,z) = {\,\int_{z'=\bar z}^z B(z,z')\over 2\,(y(z)-y(\bar z))\,dx(z)}.
\eeq
$W_n^{(g)}(z_1,\dots,z_n)$ is a meromorphic 1-form for each  $z_i\in\curve$, it is symmetric in all the $z_i$'s. For $2g-2+n>0$, it has poles only at branchpoints, without residues, and the degree of the poles are $\leq 6g-4+2n$.

$W_n^{(g)}$ with $2-2g-n<0$ are called stable, and those with $2-2g-n\geq 0$ are called unstable (only $W_1^{(0)}$ and $W_2^{(0)}$ are unstable).
\ed

The symplectic invariants themselves are $F_g=W_{n=0}^{(g)}$ for $n=0$, and are defined as follows
\bd\label{defFg}
For $g\geq 2$, the symplectic invariants of $\spcurve$ are defined by
\beq
F_g(\spcurve)=W_0^{(g)}(\spcurve) = {1\over 2-2g}\,\sum_{a={\rm branch\, points}}\Res_{z\to a}\,
W_1^{(g)}(z)\,\Phi(z)
\eeq
where $\Phi(z)$ is any function defined locally near $a$ such that
\beq
d\Phi=y\,dx.
\eeq
For $g=1$, $F_1$ is defined as
\beq
F_1(\spcurve) = {1\over 24}\,\ln{\left(\tau_B(\{{\rm x}_i=x(a_i)\, |\,a_i={\rm branch\, points}\})\,\,\prod_{a={\rm branch\, points}} y'(a)\right)}
\eeq
where $y'(a)=\lim_{z\to a}\, {y(z)-y(a)\over \sqrt{x(z)-x(a)}}$, and
$\tau_B({\rm x}_1,\dots,{\rm x}_k)$ is the Bergman Tau-function defined by \cite{Kokotov2002}
\beq
{\d \tau_B({\rm x}_1,\dots,{\rm x}_k)\over \d\,{\rm x}_i }= \Res_{z\to a_i}\,{B(z,\bar z)\over dx(z)}
\eeq
where ${\rm x}_i=x(a_i)$ are the $x$--projections of branchpoints.
There is also a definition of $F_0(\spcurve)$, see \cite{EOFg}, but we shall not use it in this article, we just notice that $F_0(\spcurve)$ doesn't depend on the Bergman kernel.
\ed

Our goal, is to relate symplectic invariants of an arbitrary spectral curve, to the combinatorics of intersection numbers in the moduli space of curves.

\subsubsection{Examples with 1 branchpoint}

Assume that there is only one branchpoint at $z=a$.
It is convenient to define:
$$
d\xi_d(z) = -\Res_{z'\to a}\,\,B(z,z')\,\,{(2d-1)!!\over 2^d\,\,(x(z')-x(a))^{d+{1\over 2}}}
$$
and the Taylor expansion of $y(z)$ near $z=a$:
\bea\nonumber
y(z)
&=&y(a) + \sum_{k=0}^\infty t_{k+2}\,(x(z)-x(a))^{k\over 2} \cr
&=& y(a)+t_3\,(x(z)-x(a))^{1\over 2}+t_4\,(x(z)-x(a))+t_5\,(x(z)-x(a))^{3\over 2}+\dots .
\eea
and the Taylor expansion of $B(z,z')$ near $a$:
\bea\nonumber
B(z,z') &=& {dx(z)\otimes dx(z')\over 4\sqrt{x(z)-x(a)}\,\sqrt{x(z')-x(a)}}\,\,\Big[\,{1\over (\sqrt{x(z)-x(a)}-\sqrt{x(z')-x(a)})^2} \cr
&& + \sum_{k,l} B_{k,l}\, (x(z)-x(a))^{k\over 2}\,(x(z)-x(a))^{l\over 2}\,dx(z)\,dx(z') \Big]
\eea

\bigskip

For low values of $g$ and $n$, a direct computation of residues gives:

$\bullet$
\beq\label{W30exspinv}
W_3^{(0)}(z_1,z_2,z_3) = {1\over 2t_3}\,d\xi_0(z_1)\,d\xi_0(z_2)\,d\xi_0(z_3)
\eeq

$\bullet$
\beq\label{W11exspinv}
W_1^{(1)}(z) = {1\over 24\,t_3}\,\left(d\xi_1(z)-{3t_5\over 2t_3}\,d\xi_0(z)\right) + {B_{0,0}\over 4t_3}\,d\xi_0(z)
\eeq

$\bullet$
\bea\label{W40exspinv}
W_4^{(0)}(z_1,z_2,z_3,z_4) 
&=& {1\over 2t_3^2}\,\left(d\xi_1(z_1)d\xi_0(z_2)d\xi_0(z_3)d\xi_0(z_4)+{\rm sym}\right)\cr
&&  -{3t_5\over 4t_3^3}\,\, d\xi_0(z_1)d\xi_0(z_2)d\xi_0(z_3)d\xi_0(z_4) \cr
&&  + {3\over 4\,t_3^2}\, B_{0,0}\,d\xi_0(z_1)d\xi_0(z_2)d\xi_0(z_3)d\xi_0(z_4)
\eea
and so on...
Our goal, is to interpret the coefficients, like $1/24 t_3$, or $-3t_5/2t_3$, or $B_{0,0}/4t_3$, in terms of intersection numbers.

\section{Intersection numbers}

\subsection{Definitions}

Let ${\cal M}_{g,n}$ be the moduli space of complex curves of genus $g$ with $n$ marked points.
It is a complex orbifold (manifold quotiented by a group of symmetries), of dimension
\beq
{\rm dim}\,{\cal M}_{g,n}=d_{g,n}=3g-3+n.
\eeq
Let $(C,p_1,\dots,p_n)\in {\cal M}_{g,n}$ be a complex curve $C$ with $n$ marked points $p_1,\dots,p_n$.
Let ${\cal L}_i$ be the cotangent bundle at $p_i$, i.e. the bundle over ${\cal M}_{g,n}$ whose fiber is the cotangent space $T^*(p_i)$ of $C$ at $p_i$.
It is customary to denote its first Chern class:
\beq
\psi_i=c_1({\cal L}_i).
\eeq
$\psi_i$ is (the cohomology equivalence class modulo exact forms, of) a 2-form on ${\cal M}_{g,n}$, therefore it makes sense to compute the "intersection number"
\beq
\left<\psi_1^{d_1}\dots \psi_n^{d_n}\right>_{g,n} = \int_{[\overline{\cal M}_{g,n}]^{\rm vir}}\psi_1^{d_1}\dots \psi_n^{d_n}
\eeq
on the compactification $\overline{\cal M}_{g,n}$ of ${\cal M}_{g,n}$ (or more precisely, on a virtual cycle $[]^{\rm vir}$ of $\overline{\cal M}_{g,n}$, taking carefully account of the non-smooth curves at the boundary of ${\cal M}_{g,n}$), provided that
\beq
\sum_i d_i=d_{g,n}=3g-3+n.
\eeq
If this equality is not satisfied we define $\left<\psi_1^{d_1}\dots \psi_n^{d_n}\right>_{g,n} =0$.

\medskip

More interesting characteristic classes and intersection numbers are defined as follows.
Let (we follow the notations of \cite{Liu2009}, and refer the reader to it for details)
$$
\pi:\overline{\cal M}_{g,n+1}\to \overline{\cal M}_{g,n}
$$
be the forgetful morphism (which forgets the last marked point), and let $\sigma_1,\dots,\sigma_n$ 
be the canonical sections of $\pi$, and $D_1,\dots,D_n$ be the corresponding divisors in $\overline{\cal M}_{g,n+1}$. Let $\om_\pi$ be the relative dualizing sheaf.
We consider the following tautological classes on $\overline{\cal M}_{g,n}$:

$\bullet$ The $\psi_i$ classes (which are 2-forms), already introduced above:
$$ \psi_i = c_1(\sigma_i^*(\om_\pi)) $$
It is customary to use Witten's notation:
\beq\label{deftaud}
\psi_i^{d_i}=\tau_{d_i}.
\eeq

$\bullet$ The Mumford $\kappa_k$ classes \cite{Mumford1983, Arbarello1996}:
$$ \kappa_k = \pi_*(c_1(\om_\pi(\sum_i D_i))^{k+1} ) .$$
$\kappa_k$ is a $2k$--form. 
$\kappa_0$ is the Euler class, and in ${\cal M}_{g,n}$, we have
$$
\kappa_0=-\chi_{g,n}=2g-2+n.
$$
$\kappa_1$ is known as the Weil-Petersson form since it is given by $2\pi^2\kappa_1=\sum_i dl_i\wedge d\theta_i$ in the Fenchel-Nielsen coordinates $(l_i,\theta_i)$ in Teichm\"uller space \cite{Wolpert1983}.

In some sense, $\kappa$ classes are the remnants of the $\psi$ classes of (clusters of) forgotten points.
There is the formula \cite{Arbarello1996}:
\beq
\pi_* \psi_1^{d_1}\dots\psi_n^{d_n}\,\psi_{n+1}^{k+1}\, = \psi_1^{d_1}\dots\psi_n^{d_n}\,\kappa_k
\eeq
\beq
\pi_*\pi_* \psi_1^{d_1}\dots\psi_n^{d_n}\,\psi_{n+1}^{k+1}\,\psi_{n+2}^{k'+1}\, = \psi_1^{d_1}\dots\psi_n^{d_n}\,(\kappa_k\,\kappa_{k'}+\kappa_{k+k'})
\eeq
and so on...

\medskip
$\bullet$ The Hodge class  $\Lambda(\alpha)=1+\sum_{k=1}^g \,(-1)^k\,\alpha^{-k}\,c_k(\mathbb E)$ where $c_k(\mathbb E)$ is the $k^{\rm th}$ Chern class of the Hodge bundle $\mathbb E=\pi_*(\om_\pi)$.
Mumford's formula \cite{Mumford1983, FaberC.1998} says that
\beq
\Lambda(\alpha)= \ee{\sum_{k\geq 1} {{B_{2k}\,\alpha^{1-2k}\over 2k(2k-1)}\,\,\left(\kappa_{2k-1}-\sum_i \psi_i^{2k-1}+{1\over 2}\sum_\delta \sum_j (-1)^j\,\,l_{\delta*} \psi^j\,\psi'^{2k-2-j}\right)}}
\eeq
where $\Ber_{k}$ is the $k^{\rm th}$ Bernoulli number, $\delta$ a boundary divisor (i.e. a cycle which can be pinched so that the pinched curve is a stable nodal curve, i.e. replacing the pinched cycle by a pair of marked points, all components have a strictly negative Euler characteristics), and $l_{\delta*}$ is the natural inclusion from  $\overline{\cal M}_{g,n}$ to $\overline{\cal M}_{g-1,n+2}+\sum'_{h,m} \overline{\cal M}_{h,m+1}\times \overline{\cal M}_{g-h,n-m+1}$, where $\sum'_{h,m}$ means that the sum is restricted to stable moduli spaces only. In other words $\sum_\delta l_{\delta*}$ adds a nodal point in all possible ways. 

\bigskip
In fact, all tautological classes in $\overline{\cal M}_{g,n}$ can be expressed in terms of $\psi$-classes or their pull back or push forward from some $\overline{\cal M}_{h,m}$ \cite{Bertram2006}.
Faber's conjecture \cite{FaberC.1998} (partly proved in \cite{Mulase2006} and \cite{Liu2009}) proposes an efficient method to compute intersection numbers of $\psi, \kappa$ and Hodge classes.

\subsection{Some already known cases}

It is already known that :
\bt 
If $\spcurve$ is the Airy curve $y=\sqrt x$, i.e. more precisely $\spcurve=(\mathbb C,x(z)=z^2,y(z)=z, B(z,z')=dz\otimes dz'/(z-z')^2)$, one has for $2g-2+n>0$
\beq
W_n^{(g)}(z_1,\dots,z_n) = (-2)^{\chi_{g,n}}\sum_{d_1+\dots+d_n= d_{g,n}} \prod_{i=1}^n\,{(2d_i+1)!!\,dz_i\over z_i^{2d_i+2}}\,\left<\prod_{i=1}^n\,\psi_i^{d_i}\right>_{g,n}.
\eeq
In other words the symplectic invariants of the Airy curve, generate intersection numbers of $\psi$ classes. For the airy spectral curve, we have $F_g=0$.
\et

This theorem is a corollary of the following one, slightly more general:
\bt[proved in \cite{Mulase2006, Liu2007, Eynard}]\label{thspinvKonts} 
If $\spcurve$ is the deformed Airy curve $y=\sum_k t_{k+2}\,x^{k/2}$, i.e. more precisely $\spcurve=(\mathbb C,x(z)=z^2,y(z)=\sum_k t_{k+2}\,z^k, B(z,z')=dz\otimes dz'/(z-z')^2)$, one has for $2g-2+n>0$
\beq
W_n^{(g)}(z_1,\dots,z_n) = (-2)^{\chi_{g,n}}\sum_{d_1+\dots+d_n\leq d_{g,n}} \prod_{i=1}^n\,{(2d_i+1)!!\,dz_i\over z_i^{2d_i+2}}\,\left<\prod_{i=1}^n\,\psi_i^{d_i}\,\,\ee{\sum_k \td t_{k} \kappa_k}\right>_{g,n}.
\eeq
In particular for $n=0$ and $g\geq 2$
\beq
F_g = 2^{2-2g}\,\,\left<\ee{\sum_k \td t_{k} \kappa_k}\right>_{g,0},
\eeq
where the dual times $\td t_k$ are defined by
\beq
\ee{-\sum_k \td t_k\,u^{-k}} = \sum_{k} {(2k+1)!!}\,\,t_{2k+3}\,u^{-k}.
\eeq
In other words the symplectic invariants of the deformed Airy curve, generate intersection numbers of $\psi$ and $\kappa$ classes.

\et

\proof{This theorem can be deduced from the work of \cite{Mulase2006, Liu2007} through Virasoro constraints. Another proof can be found in \cite{Eynard} by an argument similar to Kontsevich's \cite{Kontsevich1992}, i.e. using the Strebel decomposition of the moduli space, to write those intersection numbers as expectation values of the Kontsevich matrix integral, and then computing those expectation values by integrating by parts in the matrix integral (i.e. solving loop equations).}

\medskip
Our goal is to generalize those formulae relating symplectic invariants to intersection numbers, to arbitrary spectral curves.

\subsection{Main theorem}

Our main theorem is

\bt\label{thspinv1bp}
Let $\spcurve_a=(\curve,x,y,B)$ be a spectral curve, with only one branchpoint $a$.
Its symplectic invariant descendents, for $2-2g-n<0$, are given by the intersection numbers:
\beq\label{maineq}
\encadremath{
W_n^{(g)}(\spcurve_a;z_1,\dots,z_n) = 2^{d_{g,n}}\!\!\!\!\sum_{d_1+\dots+d_n\leq d_{g,n}}\, \prod_i d\xi_{d_i}(z_i)\,\,\left< \ee{{1\over 2}\sum_\delta l_{\delta*}\hat B(\psi,\psi')}\,\ee{\sum_k \td t_{k} \kappa_k}\prod_i \psi_i^{d_i}\right>_{{g,n}}
}\eeq
In particular for $n=0$, the symplectic invariants $F_g=W_0^{(g)}$ for $g\geq 2$ are the following intersection numbers
\beq
F_g(\spcurve_a) = 2^{3g-3}\,\,\left< \ee{{1\over 2}\sum_\delta l_{\delta*}\hat B(\psi,\psi')}\,\ee{\sum_k \td t_{k} \kappa_k}\right>_{{g,0}}.
\eeq

In this formula:

$\bullet$ the times $\td t_{k}$ are computed from the Laplace transform of the 1-form $ydx$
\beq
\ee{-\sum_k \td t_{k} u^{-k}} = {2\,u^{3/2}\,\ee{u x(a)}\over \sqrt{\pi}}\,\int_\gamma \ee{-\,ux}\,ydx
\eeq
where $\gamma$ is a steepest descent path from the branchpoint to $x=+\infty$, i.e. $x(\gamma)-x(a)=\mathbb R_+$.

$\bullet$ the 1-forms $d\xi_d(z)$ are defined by
\beq
d\xi_d(z) = -\Res_{z'\to a}\,B(z,z')\,\, {(2d-1)!!\over 2^d\,(x(z')-x(a))^{d+1/2}}
\eeq

$\bullet$ the kernel $\hat B$ 
\beq
\hat B(\psi,\psi') = \sum_{k,l} \hat B_{k,l}\,\psi^k\,\psi'^l
\eeq
is defined by the double Laplace transform of the Bergman kernel:
\beq
\sum_{k,l} \hat B_{k,l}  u^{-k}\,u'^{-l} = {(uu')^{1/2}\,\ee{(u+u')\,x(a)}\over 2\pi}\,\int_{z\in\gamma}\int_{z'\in\gamma} \ee{-\,ux(z)}\,\ee{-\,u'x(z')}\,\left(B(z,z')-\Brond(z_1,z_2)\right)
\eeq
where the integral is regularized by substracting the "trivial part" of the double pole
\beq
\Brond(z_1,z_2) = {dx(z_1)\otimes dx(z_2)\over 4\sqrt{x(z_1)-x(a)}\,\sqrt{x(z_2)-x(a)}}\,\,{1\over (\sqrt{x(z_1)-x(a)}-\sqrt{x(z_2)-x(a)})^2}.
\eeq

$\bullet$ $\sum_\delta$ means the sum over all boundary divisors $\delta$, and $l_{\delta}*$ is the "operator pinching the boundary cycle $\delta$" to a nodal point. It adds a nodal point, i.e. two marked points, respecting stability constraints (each component must be stable, i.e. have strictly negative Euler characteristics).
$l_{\delta*}$ is the natural inclusion from  $\delta{\cal M}_{g,n}$ to ${\cal M}_{g-1,n+2}+\sum'_{h,m} {\cal M}_{h,m+1}\times {\cal M}_{g-h,n-m+1}$, where $\sum'$ means that the sum is restricted to stable moduli spaces only.
$\psi$ and $\psi'$ are the first Chern classes of the cotangent line bundle of the nodal point.

\medskip
Written in Laplace transform, \eq{maineq} reads 
\bea
&& \prod_{i=1}^n \sqrt{\mu_i\over \pi}\,\ee{\mu_i\,x(a)}\,\int_{z_1\in\gamma}\dots\int_{z_n\in\gamma} 
\prod_{i=1}^n \ee{-\mu_i x(z_i)}\,\,W_n^{(g)}(z_1,\dots,z_n) \cr
&=& 2^{d_{g,n}+n} \left< \prod_{i=1}^n \hat B(\mu_i,1/\psi_i)\,\, \ee{{1\over 2}\sum_\delta \sum_{k,l} l_{\delta*}\psi^k\,\psi'^l}\,\,\ee{\sum_k \td t_k \kappa_k}\right>_{g,n}
\eea
where
\bea\label{eqdefBhatuv}
\hat B(u,v) 
&=&  {(uv)^{1/2}\,\ee{(u+v)\,x(a)}\over 2\pi}\,\int_{z\in\gamma}\int_{z'\in\gamma} \ee{-\,ux(z)}\,\ee{-\,vx(z')}\,B(z,z') \cr
&=& {uv\over u+v} + \sum_{k,l} \hat B_{k,l}\,u^{-k}\,v^{-l} \cr
&=& \sum_k (-1)^{k} u^{k+1}\,v^{-k} + \sum_{k,l} \hat B_{k,l}\,u^{-k}\,v^{-l} .\cr
\eea
\et

We shall prove this theorem below in section \ref{secproofmainth} of this article.

Before, let us see some applications.

\subsection{How to use the formula}

Let us show how to use the formula of theorem \ref{thspinv1bp}.
First, one needs to know that $\psi$ is a 2-form, it is assigned a degree $1$, and $\kappa_k$ is a $2k$--form, which is assigned  degree $k$.
$\kappa_0=-\chi=2g-2+n$ is a number (degree $0$), and can be factored out of the intersection number:
$$
\ee{\td t_0\kappa_0} \to \ee{(2g-2+n)\td t_0} = (2t_3)^{2-2g-n}.
$$

An intersection number is non--zero only if the total degree is $d_{g,n}=3g-3+n$. This means the $\ee{\sum_k \td t_k \kappa_k}$ can be truncated to $k\leq 3g-3+n$, and the exponential can be Taylor expanded and the Taylor expansion can be truncated to order $\leq 3g-3+n$.

\medskip

Similarly, notice that $l_*$ diminishes $d_{g,n}$ by $1$, so we may truncate the Taylor expansion of $\ee{{\hat B_{k,l}\over 2}\,l^*\psi^k\psi'^l}$ to order $d_{g,n}$:
\beq
\ee{\sum_{k,l}{\hat B_{k,l}\over 2}\,l^*\psi^k\psi'^l} \longrightarrow 1+\sum_{j=1}^{d_{g,n}}\,{1\over j!}\,\sum_{k_1,\dots,k_j,l_1,\dots,l_j}\, \prod_{i=1}^j {\hat B_{k_i,l_i}\over 2}\,l^*\psi_{n+2i-1}^{k_i}\psi_{n+2i}^{l_i}.
\eeq

\medskip
$\bullet$ For example for $g=0, n=3$, we have $d_{0,3}=0$ and $\kappa_0=-\chi_{0,3}=1$, so that we may replace in ${\cal M}_{0,3}$
\beq
\ee{\sum_k \td t_k \kappa_k} \longrightarrow \ee{\td t_0}\
\virg
\ee{\sum_{k,l}{\hat B_{k,l}\over 2}\,l^*\psi^k\psi'^l} \longrightarrow 1.
\eeq
We thus have
\beq
W_3^{(0)}(z_1,z_2,z_3) = \ee{\td t_0}\,d\xi_0(z_1)\,d\xi_0(z_2)\,d\xi_0(z_3) = {1\over 2 t_3}\,d\xi_0(z_1)\,d\xi_0(z_2)\,d\xi_0(z_3)
\eeq
which agrees with \eq{W30exspinv}

\smallskip
$\bullet$ For example for $g=1, n=1$, we have $d_{1,1}=1$ and $\kappa_0=-\chi_{1,1}=1$, so that we may replace in ${\cal M}_{1,1}$
\beq
\ee{\sum_k \td t_k \kappa_k} \longrightarrow \ee{\td t_0}\,\,(1+\td t_1 \kappa_1).
\eeq
Also, since $d_{1,1}=1$, we can replace
\beq
\ee{{\hat B_{k,l}\over 2}\,l^*\psi^k\psi'^l} \longrightarrow 1+{\hat B_{k,l}\over 2}\,l^*\psi^k\psi'^l.
\eeq
The boundary of ${\cal M}_{1,1}$ is a single point, identified with ${\cal M}_{0,3}$. Indeed, there is only one possibility of pinching a cycle for a curve in ${\cal M}_{1,1}$, i.e. the torus degenerates into a sphere with 1 nodal point, i.e. a sphere with 3 marked points in ${\cal M}_{0,3}$.
We have
\beq
<(l_* \psi^k\psi'^l)\,\,\ee{\sum_j \td t_j \kappa_j}\,\psi_1^{d_1}>_{1,1} = <\psi_2^k\psi_3'^l\,\,\ee{\sum_j \td t_j \kappa_j}\,\psi_1^{d_1}>_{0,3}.
\eeq
And in ${\cal M}_{0,3}$, we can replace
\beq
\ee{\sum_k \td t_k \kappa_k} \longrightarrow \ee{\td t_0}.
\eeq

Therefore, for $W_1^{(1)}$, theorem \ref{thspinv1bp} says that:
\bea
{1\over 2}W_1^{(1)}(z) 
&=& d\xi_1(z)\,\left<\psi\,\ee{\td t_0 \kappa_0}\right>_{1,1}
+ \td t_1\,d\xi_0(z)\,\left<\kappa_1\,\ee{\td t_0 \kappa_0}\right>_{1,1} \cr
&& + {\hat B_{0,0}\over 2}\,\,d\xi_0(z)\,\left<\ee{\td t_0 \kappa_0} \psi_1^0\psi^0\psi'^0\right>_{3,0} \cr
&=& \ee{\td t_0}\,\Big[d\xi_1(z)\,\left<\psi\right>_{1,1}
+ \td t_1\,\,d\xi_0(z)\,\left<\kappa_1\right>_{1,1}
+ {\hat B_{0,0}\over 2}\,\,d\xi_0(z)\,\left<\tau_0^3\right>_{3,0} \Big]\cr
&=& \ee{\td t_0}\,\Big[\,{1\over 24}\,d\xi_1(z)
+ {\td t_1\over 24}\,\,d\xi_0(z)
+ {\hat B_{0,0}\over 2}\,\,\,d\xi_0(z) \Big]\cr
\eea
where we have used $<\kappa_1>_{1,1}=1/24$ and $<\psi>_{1,1}=1/24$  (see appendix \ref{appkappaclasses}).
We have $\hat B_{0,0}=B_{0,0}/2$ and $\td t_1 = - 3 t_5/2t_3$, so that this expression agrees with the direct computation of \eq{W11exspinv}.

\smallskip
$\bullet$ For example for $g=0, n=4$, theorem \ref{thspinv1bp} says that:
\bea
{1\over 2}W_4^{(0)}(z_1,z_2,z_3,z_4) 
&=& d\xi_1(z_1)d\xi_0(z_2)d\xi_0(z_3)d\xi_0(z_4)\,\left<\psi\,\ee{\td t_0 \kappa_0}\right>_{0,4} + {\rm sym}\cr
&& + \td t_1\,d\xi_0(z_1)d\xi_0(z_2)d\xi_0(z_3)d\xi_0(z_4)\,\,\left<\kappa_1\,\ee{\td t_0 \kappa_0}\right>_{0,4}\cr
&& + \sum_{k,l}\,{\hat B_{k,l}\over 2}\,d\xi_0(z_1)d\xi_0(z_2)d\xi_0(z_3)d\xi_0(z_4)\,\,\left<\psi_1^{0}\psi_2^{0}\psi^k\,\ee{\td t_0 \kappa_0}\right>_{0,3}\,\left<\psi_3^{0}\psi_4^{0}\psi'^l\,\ee{\td t_0 \kappa_0}\right>_{0,3}\cr
&& + {\rm sym} \cr
&=& \ee{2\td t_0}\,\,d\xi_1(z_1)d\xi_0(z_2)d\xi_0(z_3)d\xi_0(z_4) + {\rm sym}\cr
&& + \td t_1\,\ee{2\td t_0}\,d\xi_0(z_1)d\xi_0(z_2)d\xi_0(z_3)d\xi_0(z_4) \cr
&& + {6\,\hat B_{0,0}\over 2}\,\ee{2\td t_0}\,d\xi_0(z_1)d\xi_0(z_2)d\xi_0(z_3)d\xi_0(z_4)
\eea
Again, this agrees with the direct computation of \eq{W40exspinv}.

\smallskip
$\bullet$  Similarly, for $W_{2}^{(1)}$, we have $d_{1,2}=2$ and $\kappa_0=-\chi_{1,2}=2$, and thus
\bea
{1\over 4}W_2^{(1)}(z_1,z_2) 
&=& (d\xi_2(z_1)d\xi_0(z_2)+d\xi_0(z_1)d\xi_2(z_2))\,\left<\psi^2\,\ee{\td t_0 \kappa_0}\right>_{1,2} 
+ d\xi_1(z_1)d\xi_1(z_2)\,\left<\psi_1\psi_2\,\ee{\td t_0 \kappa_0}\right>_{1,2} \cr
&& + \td t_1\, (d\xi_1(z_1)d\xi_0(z_2)+d\xi_0(z_1)d\xi_1(z_2))\,\left<\psi\,\ee{\td t_0 \kappa_0}\,\kappa_1\right>_{1,2} \cr
&&+ \td t_2\,d\xi_0(z_1)d\xi_0(z_2)\,\left<\,\ee{\td t_0 \kappa_0}\,\kappa_2\right>_{1,2} 
+ {1\over 2}\,\td t_1^2\,d\xi_0(z_1)d\xi_0(z_2)\,\left<\,\ee{\td t_0 \kappa_0}\,\kappa_1^2\right>_{1,2} \cr
&&+ {\hat B_{0,0}\over 2}\,\,(d\xi_1(z_1)d\xi_0(z_2)+d\xi_0(z_1)d\xi_1(z_2))\,\left<\psi\,\ee{\td t_0 \kappa_0}\,\right>_{0,4} \cr
&&+ {\hat B_{0,0}\,\td t_1\over 2}\,\,d\xi_0(z_1)d\xi_0(z_2)\,\left<\,\ee{\td t_0 \kappa_0}\,\kappa_1\right>_{0,4} + {\hat B_{1,0}}\,\,d\xi_0(z_1)d\xi_0(z_2)\,\left<\psi\,\ee{\td t_0 \kappa_0}\right>_{0,4} \cr
&& + \hat B_{0,0} (d\xi_1(z_1)d\xi_0(z_2)+d\xi_0(z_1)d\xi_1(z_2)) \left<\ee{\td t_0 \kappa_0}\right>_{0,3}\left<\psi\ee{\td t_0 \kappa_0}\right>_{1,1} \cr
&& + \hat B_{1,0} d\xi_0(z_1)d\xi_0(z_2) \left<\ee{\td t_0 \kappa_0}\right>_{0,3}\left<\psi\ee{\td t_0 \kappa_0}\right>_{1,1} \cr
&& + \hat B_{0,0}\td t_1\,\, d\xi_0(z_1)d\xi_0(z_2) \left<\ee{\td t_0 \kappa_0}\right>_{0,3}\left<\ee{\td t_0 \kappa_0}\,\kappa_1\right>_{1,1} \cr
&& + 6\,{\hat B_{0,0}^2\over 8}\,\, d\xi_0(z_1)d\xi_0(z_2) \left<\ee{\td t_0 \kappa_0}\right>_{0,3}\left<\ee{\td t_0 \kappa_0}\,\right>_{0,3} \cr
&& + 2\,{\hat B_{0,0}^2\over 8}\,\, d\xi_0(z_1)d\xi_0(z_2) \left<\ee{\td t_0 \kappa_0}\right>_{0,3}\left<\ee{\td t_0 \kappa_0}\,\right>_{0,3} \cr
\eea
Namely:
\bea
W_2^{(1)}(z_1,z_2) 
&=& 4\,\ee{2\td t_0}\Big[ {1\over 24}\,(d\xi_2(z_1)d\xi_0(z_2)+d\xi_0(z_1)d\xi_2(z_2))
+ {1\over 24}d\xi_1(z_1)d\xi_1(z_2) \cr
&& + {\td t_1\over 12}\, (d\xi_1(z_1)d\xi_0(z_2)+d\xi_0(z_1)d\xi_1(z_2)) \cr
&&+ {\td t_2\over 24}\,d\xi_0(z_1)d\xi_0(z_2)\,
+ \,{\td t_1^2\over 16}\,d\xi_0(z_1)d\xi_0(z_2) \cr
&&+ {\hat B_{0,0}\over 2}\,\,(d\xi_1(z_1)d\xi_0(z_2)+d\xi_0(z_1)d\xi_1(z_2)) \cr
&&+ {\hat B_{0,0}\,\td t_1\over 2}\,\,d\xi_0(z_1)d\xi_0(z_2) + {\hat B_{1,0}}\,\,d\xi_0(z_1)d\xi_0(z_2) \cr
&& + {\hat B_{0,0}\over 24} (d\xi_1(z_1)d\xi_0(z_2)+d\xi_0(z_1)d\xi_1(z_2))  \cr
&& + {\hat B_{1,0}\over 24} d\xi_0(z_1)d\xi_0(z_2)   + {\hat B_{0,0}\td t_1\over 24}\,\, d\xi_0(z_1)d\xi_0(z_2)   + \hat B_{0,0}^2\,\, d\xi_0(z_1)d\xi_0(z_2) \Big] \cr
\eea

\smallskip
$\bullet$  For example for $g=2, n=0$, we have $d_{2,0}=3$ and $\chi_{2,0}=-2$, and theorem \ref{thspinv1bp} says that:
\bea
{1\over 8}F_2 
&=& \left<\ee{\sum_k \td t_k \kappa_k} \right>_{2,0}  \cr
&& +{1\over 2}\sum_{i+j\leq 2}\, {\hat B_{i,j}} \Big[ \left<\psi_1^i\psi_2^j\ee{\sum_k \td t_k \kappa_k}\right>_{1,2}   + \left<\psi_1^i\ee{\sum_k \td t_k \kappa_k}\right>_{1,1}\,\left<\psi_2^j\ee{\sum_k \td t_k \kappa_k}\right>_{1,1} \Big] \cr  
&& +{1\over 8}\sum_{i+j+m+n\leq 1}\,\hat B_{i,j}\hat B_{m,n} \Big[ \left<\psi_1^i\psi_2^j\psi_3^m\psi_4^n\,\ee{\sum_k \td t_k \kappa_k} \right>_{0,4} \cr
&& + 2\, \left<\psi_1^i\psi_2^j\psi_3^m\,\ee{\sum_k \td t_k \kappa_k} \right>_{0,3}\,\left<\psi_4^n\,\ee{\sum_k \td t_k \kappa_k} \right>_{1,1}  \cr
&& +2\, \left<\psi_1^i\psi_3^m\psi_4^n\,\ee{\sum_k \td t_k \kappa_k} \right>_{0,3}\,\left<\psi_2^j\,\ee{\sum_k \td t_k \kappa_k} \right>_{1,1} \Big] \cr
&& +{1\over 48}\sum_{i+j+m+n+p+q\leq 0}\,\hat B_{i,j}\hat B_{m,n}\hat B_{p,q}\,\Big[ \cr
&& 2\, \left<\psi_1^i\psi_2^j\psi_5^p\,\ee{\sum_k \td t_k \kappa_k} \right>_{0,3}\,\left<\psi_3^m\psi_4^n\psi_6^q\,\ee{\sum_k \td t_k \kappa_k} \right>_{0,3} \cr
&& 
+2\,\left<\psi_1^i\psi_3^m\psi_5^p\,\ee{\sum_k \td t_k \kappa_k} \right>_{0,3}\,\left<\psi_2^j\psi_4^n\psi_6^q\,\ee{\sum_k \td t_k \kappa_k} \right>_{0,3} \cr
&&
+2\,\left<\psi_1^i\psi_4^n\psi_5^p\,\ee{\sum_k \td t_k \kappa_k} \right>_{0,3}\,\left<\psi_2^j\psi_3^m\psi_6^q\,\ee{\sum_k \td t_k \kappa_k} \right>_{0,3}  \cr
&& + 2\left<\psi_1^i\psi_2^j\psi_3^m\,\ee{\sum_k \td t_k \kappa_k} \right>_{0,3}\,\left<\psi_4^n\psi_5^p\psi_6^q\,\ee{\sum_k \td t_k \kappa_k} \right>_{0,3} \cr
&& + 2\,\left<\psi_1^i\psi_3^m\psi_4^m\,\ee{\sum_k \td t_k \kappa_k} \right>_{0,3}\,\left<\psi_2^j\psi_5^p\psi_6^q\,\ee{\sum_k \td t_k \kappa_k} \right>_{0,3} \Big]\cr
\eea
Namely, that gives (see appendix \ref{appkappaclasses})
\bea
F_2
&=& 8\ee{2\td t_0}\,\Big\{
{\td t_3\over 3^2\,\,2^7}+{\td t_2 \td t_1\over 15\,\,2^4}+{43\,\td t_1^3\over 5\,\, 3^3\,\,2^7}
+{\hat B_{0,0}\over 2}\left( {\td t_2\over 24}+{\td t_1^2\over 16}\right) + {\hat B_{1,0}\,\td t_1\over 12}
+ {\hat B_{1,1}\over 48}+ {\hat B_{2,0}\over 48}\cr
&& + {\hat B_{0,0}\,\td t_1^2\over 2*24^2}+ {\hat B_{1,0}\,\td t_1\over 24^2}+ {\hat B_{1,1}\,\over 2*24^2}
+ {\hat B_{0,0}^2\td t_1\over 8}\,\left(1+{4\over 24}\right)+ {\hat B_{0,0}\,\hat B_{1,0}\over 8}\,\left(4+{4\over 24}\right) \cr
&& + {10\,\hat B_{0,0}^3\over 48}\, \,\Big\}.
\eea
One can check that this agrees with the direct computation of symplectic invariants, computing the residues in def. \ref{defFg}.

\section{The topological vertex}\label{secvertex}

Specializing theorem \ref{thspinv1bp} to the topological vertex's \cite{Aganagic2004, KatzSheldon2001, LiJun2004} spectral curve we get:

\bt[Topological vertex and BKMP]\label{cortopvertex}
For any framing $f$, choose the framed topological vertex spectral curve
 $\spcurve_{\rm vertex}=(\mathbb C^*\setminus ]-\infty,0]\cup[1,\infty[, x(z)=-f \ln{z}-\ln{(1-z)},y(z)=-\ln{z},B(z,z')=dzdz'/(z-z')^2)$, i.e.:
$$
\ee{-x}=\ee{-fy}\,(1-\ee{-y}).
$$
Then we have
\beq
W_n^{(g)}(\spcurve_{\rm vertex};z_1,\dots,z_n) = 2^{d_{g,n}}\!\!\!\!\! \sum_{d_1+\dots+d_n\leq 3g-3+n}\, \prod_i d\hat\xi_{d_i}(z_i)\,\,\left< \Lambda(1)\Lambda(f)\Lambda(-1-f)\,\,\prod_i \psi_i^{d_i}\right>_{{g,n}}
\eeq
where $\Lambda(\alpha)=1+\sum_{k=1}^g \,(-1)^k\,\alpha^{-k}\,c_k(\mathbb E)$ is the Hodge class ($c_k(\mathbb E)$ is the $k^{\rm th}$ Chern class of the Hodge bundle $\mathbb E=\pi_*(\om_\pi)$), and
\beq
\xi_d(z) = {f+1\over f}\,\sum_{\mu=0}^\infty \ee{-\mu x(z)}\,{(-\mu)^{d}\,(\mu(1+1/f))!\over \mu!\,(\mu/ f)!}
\eeq
In other words we recognize Mari\~no--Vafa formula \cite{MV01} for the topological vertex
\bea
W_n^{(g)}(\spcurve_{\rm vertex};z_1,\dots,z_n) 
&=& \sum_{\mu_1,\dots,\mu_n} \prod_i {(\mu_i(f+1))!\over \mu_i!\,(f\mu_i)!}\,\,\ee{-\mu_i x(z_i)}\mu_i dx(z_i)\, \cr
&& \,\,\left< {\Lambda(1)\Lambda(f)\Lambda(-1-f)\over\prod_i  1+\mu_i\psi_i}\right>_{{g,n}}.
\eea
This gives a new proof of the BKMP conjecture \cite{BKMP} for the vertex (already proved in \cite{ChenLin2009, ZhouJian2003}), i.e. that the symplectic invariants of the framed vertex spectral curve $\spcurve_{\rm vertex}$ (which is the mirror curve of $\mathbb C^3$), are the Gromov-Witten invariants of the framed vertex.

\et

\proof{We prove this theorem in section \ref{secproofvertex} below. We just mention that this theorem is already known from \cite{ChenLin2009, ZhouJian2003}. We just propose a new proof using only the topological recursion.}


\section{Several branchpoints}

Theorem \ref{thspinv1bp} Immediately generalizes to several branchpoints.

\bt\label{thspinvmultibp}
Let $\spcurve=(\curve,x,y,B)$ be a spectral curve with branchpoints $a_1,\dots,a_\ell$.
Its symplectic invariant descendents of $\spcurve$ are given by
\bea
W_n^{(g)}(\spcurve;z_1,\dots,z_n)
&=& 2^{d_{g,n}}\,\sum_{d_1+\dots+d_n\leq d_{g,n}}\,\sum_m\qquad\sum_{{\rm deg.}\,{\cal M}_{g,n}\to \cup_{j=1}^m {\cal M}_{g_j,n_j+k_j}} \quad \sum_{1\leq a_j\leq \ell,\, j=1,\dots,m} \cr
&& \sum_{I_1\uplus\dots\uplus I_m=J,\, \#I_j=n_j} \quad\sum_{d_{j,i},\,i=1,\dots,k_j} {1\over \#{\rm Aut}}\,\prod_{j<j'} \prod_{i=1}^{k_j}\prod_{i'=1}^{k_{j'}}\, \hat B_{a_j,d_{j,i};a_{j'},d_{j',i'}}\cr
&& \prod_{j=1}^{m} \prod_{z_i\in I_j} d\xi_{a_j,d_{i}}(z_i)\,\left< \Lambda_{a_j}\,\,\prod_{z_i\in I_j}\tau_{d_i}\, \prod_{i=1}^{k_j} \tau_{d_{j,i}}\right>_{g_j,n_j+k_j}
\eea
where $J=\{z_1,\dots,z_n\}$ and we sum over all stable degeneracies of ${\cal M}_{g,n}$ made of $m$ stable connected components, the $j^{\rm th}$ component having genus $g_j$, having $n_j$ marked points, and $k_j$ nodal points.

We have defined

$\bullet$ The forms $d\xi_{a,d}(z)$ for each branchpoint $a$
\beq
d\xi_{a,d}(z) = -\,\Res_{z'\to a}\,B(z,z')\,{(2d-1)!!\over 2^d\,(x(z')-x(a))^{d+1/2}}
\eeq

$\bullet$ The double Laplace transfoms of the Bergman kernel
\bea
&& \sum_{k,k'}\,\hat B_{a,k;a',k'}\,u^{-k}\,v^{-k'} \cr
&=& (1-\delta_{a,a'})\,{\sqrt{uv}\over 2\pi}\int_{z\in\gamma_a} \ee{-u(x(z)-x(a))}\,\int_{z'\in\gamma_{a'}} \ee{-v(x(z')-x(a'))}\,B(z,z')
\eea

$\bullet$ The tautological class $\L_a$ associated to the branchpoint $a$:
\beq
\Lambda_a = \ee{\sum_k \td t_{a,k}\kappa_k}\,\,\ee{{1\over 2}\sum_\delta\sum_{k,l} \hat B_{a,k;a,l} l_{\delta*} \psi^k\psi'^l}
\eeq
where $\td t_k$ are the dual times
\beq
\ee{-\sum_k \td t_{a,k}\,u^{-k}} = {2\,\sqrt u\,\,\ee{ux(a)}\over \sqrt\pi}\,\int_{\gamma_a} \ee{-ux(z)}\,dy(z)
\eeq
where the steepest descent contour $\gamma_a$ for a branchpoint $a$, is a connected arc on $\curve$, going through $a$, and such that $x(\gamma_a)-x(a)=\mathbb R_+$.
And the $\hat B_{a,k,k'}$ are given by
\bea
&& \sum_{k,k'}\,\hat B_{a,k,k'}\,u^{-k}\,v^{-k'} \cr
&=& {\sqrt{uv}\over \pi}\int_{z\in\gamma_a} \ee{-u(x(z)-x(a))}\,\int_{z'\in\gamma_{a}} \ee{-v(x(z')-x(a))}\,B(z,z')-\Brond_a(z,z')).
\eea

\et

\proof{This theorem is the immediate generalization of theorem \ref{thspinv1bp}, using the methods of \cite{OrantinN.2008, Kostov2010}, or an immediate generalization of lemma \ref{mainlemma} poved in appendix \ref{applemma}.}

\section{Proof of the main theorem}\label{secproofmainth}

Let us prove theorem \ref{thspinv1bp}.

\subsection{Kontsevich's curve Symplectic invariants}

Consider a spectral curve $\spcurve_a=(\curve,x,y,B)$ where $\curve$ contains only one branch-point located at $a$.
Locally we write the Taylor expansion near $x=x(a)$ as:
\beq
y \mathop{\sim}_{a} \sum_k t_{k+2}\,\,(x-x(a))^{k\over 2}.
\eeq
The Bergman kernel $B(z_1,z_2)$ is used to define symplectic invariants.
The Bergman kernel is a symmetric 2-form on $\spcurve_a\times \spcurve_a$ with a double pole on the diagonal:
\beq
B(z_1,z_2) \sim {dz_1\otimes dz_2\over (z_1-z_2)^2} + {\rm regular}
\eeq
where $z$ may be any local parameter on $\spcurve_a$, in particular, if we choose $z=\zeta=\sqrt{x-x(a)}$ near the branchpoint $a$,
and denote
\bea
\Brond(z_1,z_2) 
&=& {1\over 4\,\sqrt{(x_1-x(a))(x_2-x(a))}}\,\,{dx_1\,\otimes dx_2\over (\sqrt{x_1-x(a)}-\sqrt{x_2-x(a)})^2} \cr
&=&  {d\zeta(z_1)\otimes d\zeta(z_2)\over (\zeta(z_1)-\zeta(z_2))^2} ,
\eea
we have that
\beq
B(z_1,z_2)-\Brond(z_1,z_2) ={\rm analytical\, near\,}z_1\to a,\, z_2\to a.
\eeq

Let us now consider the same spectral curve with the Bergman kernel $\Brond$.
\beq
\spcurverond_a=(\curve,x,y,\Brond).
\eeq

$\spcurverond_a$ is the Kontsevich's spectral curve, and thus according to theorem \ref{thspinvKonts} (proved for instance in \cite{Eynard}),
the symplectic invariants $\Wrond^{(g)}_n=W_n^{(g)}(\spcurverond_a)$, are (with $\zeta_i=\sqrt{x(z_i)-x(a)}\,$):
\beq
\Wrond_n^{(g)}(\zeta_1,\dots,\zeta_n)
= (-1)^n\,2^{d_{g,n}}\,\,\sum_{d_1+\dots+d_n\leq d_{g,n}}\,\, \prod_{i=1}^n {(2d_i+1)!!\,\,d\zeta_i\over  2^{d_i}\,\zeta_i^{2d_i+2}}\,\, \left< \ee{\sum_k \td t_k \kappa_k}\,\,\prod_{i=1}^n \psi_i^{d_i}\right>_{{g,n}}
\eeq
where the times $\td t_k$ are the Schur transforms of the $t_k$'s, defined through their generating function:
\beq
\ee{-\sum_{k=0}^\infty \td t_k\, u^{-k}} = 2\sum_{k=0}^\infty {(2k+1)!!\over 2^k}\, t_{2k+3}\, \,u^{-k}.
\eeq

\subsection{Laplace transform and the spectral curve class}

We thus see that we are led to associate to any spectral curve $\spcurve$ with one branchpoint $a$, the following tautological class
\beq
 \ee{\sum_{k=0}^\infty \td t_k \kappa_k} = \ee{\td t_0 \kappa_0}\,\Big(1+\td t_1 \kappa_1 + ( {\td t_1^2\over 2}\kappa_1^2+\td t_2 \kappa_2 ) + (\td t_3 \kappa_3+\td t_1 \td t_2 \kappa_1 \kappa_2+{\td t^3_1\over 6}\kappa_1^3)+ \dots \Big)
\eeq
where the times $\td t_k$ are determined by the generating function
\beq
G(u) = \ee{-g(u)} = \ee{-\sum_{k=0}^\infty \td t_k\,u^{-k}} =  2\sum_{k=0}^\infty (2k+1)!!\,\,t_{2k+3}\,2^{-k}\,u^{-k}.
\eeq

Notice that we have:
\bea
\int_{x=x(a)}^\infty \ee{-\,u\, x} (y-\bar y)dx
&=& 2\,\ee{-\,u\,x(a)}\,\sum_{k=0}^\infty \int_{x(a)}^\infty t_{2k+3}\,(x-x(a))^{k+1/2}\,\ee{-\,u\,(x-x(a))}\,dx \cr
&=& 2\,\ee{-\,u\,x(a)}\,\sum_{k=0}^\infty t_{2k+3}\,\int_{0}^\infty \,\zeta^{2k+1}\,\ee{-\,u\,\zeta^2}\,2\zeta\,d\zeta \cr
&=& 2\,\ee{-\,u\,x(a)}\,\sum_{k=0}^\infty t_{2k+3}\,\int_{-\infty}^\infty \,\zeta^{2k+2}\,\ee{-\,u\,\zeta^2}\,\,d\zeta \cr
&=& 2\,\ee{-\,u\,x(a)}\,\sum_{k=0}^\infty t_{2k+3}\,(2k+1)!!\,2^{-k-1}\, u^{-k-1}\,\sqrt{\pi\over u} \cr
&=&  {1\over 2}\,\sqrt{\pi}\,u^{-3/2}\,\ee{-\,u\,x(a)}\,\, \ee{-g(u)}.
\eea
In other words, the $G(u)=\ee{-g(u)}$ function is related to the Laplace transform of $ydx$ along a contour passing through the branchpoint:
\beq
G(u)=\ee{-g(u)} =  2\,{u^{3/2}\,\ee{u\,x(a)}\over \sqrt{\pi}}\,\, \int_{\gamma_a}\, \ee{-\,u x}\,y dx
\eeq
Here, $\gamma_a$ is a contour on the spectral curve, passing through the branchpoint $a$, and whose $x$ projection is:
\beq
x(\gamma_a)=[x(a),+\infty[
\eeq

Let us also integrate by parts:
\beq
G(u)=\ee{-g(u)} = -\,2\, {u^{3/2}\,\ee{u\,x(a)}\over \sqrt{\pi}}\,\, {1\over u}\,\int_{\gamma_a}\, y\,d(\ee{-\,u x})
\eeq
i.e.
\beq\label{eqGuinteuxdy}
G(u)=\ee{-g(u)} =  2\,{u^{1/2}\,\ee{\,u\,x(a)}\over \sqrt{\pi}}\,\, \int_{\gamma_a}\, \ee{-\,u x}\,\, dy.
\eeq



\subsection{Introducing the Bergman kernel}

So far, we have computed $\Wrond_n^{(g)}$ with the Bergman kernel $\Brond$, and not with the proper Bergman kernel $B$ of the spectral curve $\spcurve$.
we now need to reintroduce the correct Bergman kernel.

First, using the local variable $\zeta(z)=\sqrt{x(z)-x(a)}$, compute the Taylor expansion of $B(z_1,z_2)$ near $a$:
\beq\label{defBkl}
B(z_1,z_2) -  \Brond(z_1,z_2) 
= \sum_{k,l} B_{k,l}\,\zeta(z_1)^k\,\zeta(z_2)^l\,\, d\zeta(z_1)\,d\zeta(z_2)  .
\eeq

\subsection{The basis $d\xi_d$}

We introduce the differential forms:
\beq
d\xi_d(z)  = -\,{(2d-1)!!\over 2^d}\,\,\Res_{z'\to a} B(z,z')\,{1\over \zeta(z')^{2d+1}}.
\eeq
They are defined globally on the Riemann surface ${\cal C}$, and they have poles only at the branch point.
We also introduce the even forms
\beq
d\td\xi_d(z)  = -\,\Res_{z'\to a} B(z,z')\,{1\over \zeta(z')^{2d}}.
\eeq

In the vicinity of the branchpoint we have the Laurent series expansion, using \eq{defBkl}:
\beq
d\xi_d(z)  = -\,{(2d+1)!!\,\,d\zeta(z)\over 2^d\,\,\zeta(z)^{2d+2}}-{(2d-1)!!\over 2^d}\,\sum_{k}\,B_{2d,k} \,\zeta(z)^k\,d\zeta(z) 
\eeq
i.e.
\beq\label{defdxidBkl}
\xi_d(z) = {(2d-1)!!\over 2^d}\,\left({1\over \zeta(z)^{2d+1}}-\sum_{k}\,B_{2d,k} {\zeta(z)^{k+1}\over k+1}\right).
\eeq
And similarly
\beq
d\td\xi_d(z)  = -2d\,{\,\,d\zeta(z)\over \zeta(z)^{2d+1}}-\,\sum_{k}\,B_{2d-1,k} \,\zeta(z)^k\,d\zeta(z) .
\eeq

They are such that when $z_1$ is in the vicinity of the branchpoint (but not necessarily $z_2$):
\bea
B(z_1,z_2) &=&\, -\,\sum_{d=0}^\infty \zeta(z_1)^{2d}\, d\zeta(z_1)\,{2^d\over (2d-1)!!}\, d\xi_d(z_2) \cr
&& - \sum_{d=0}^\infty \zeta(z_1)^{2d-1}\, d\zeta(z_1)\,\, d\td\xi_d(z_2).
\eea

Since all $W_n^{(g)}$'s are computed by taking residues at the branchpoint, we always need to replace $B(z_1,z_2)$ by its Taylor expansion, and therefore, $W_n^{(g)}$ is a linear combination of the $d\xi_{d_i}(z_i)$ and $d\td\xi_{d_i}(z_i)$.
Moreover, it is a known property (see \cite{EOFg}) of $W_n^{(g)}$ that
\beq
W_n^{(g)}(z_1,z_2,\dots,z_n)+W_n^{(g)}(\bar z_1,z_2,\dots,z_n)
\eeq
is analytical when $z_1$ is at the branchpoint, i.e. there can be only odd degree poles in $\zeta(z_1)$, and thus the even $d\td\xi_{d_i}(z_i)$ don't appear in $W_n^{(g)}$.

Therefore $W_n^{(g)}$ can be decomposed uniquely on that basis, as:
\beq
W_n^{(g)}(z_1,\dots,z_n) = 2^{d_{g,n}}\,\sum_{d_1+\dots+d_n\leq 3g-3+n}\,\, A^{(g)}_n(d_1,\dots,d_n)\,\,\prod_{i=1}^n d\xi_{d_i}(z_i).
\eeq



\medskip
We have
\beq
\Res_{z\to a} \zeta(z)^{2k+1}\,d\xi_d(z) = - (2d+1)!!\,2^{-d}\,\delta_{k,d},
\eeq
so that
\beq
2^{d_{g,n}}\,A^{(g)}_n(d_1,\dots,d_n) = (-1)^n\,\Res_{z_i\to a}\, W_n^{(g)}(z_1,\dots,z_n)  \,\,\prod_{i=1}^n {2^{d_i}\,\zeta(z_i)^{2d_i+1}\over (2d_i+1)!!}.
\eeq
In particular for Kontsevich integral, i.e. with $B=\Brond$ we have
\beq
\stackrel{{\circ}}{A}^{(g)}_n(d_1,\dots,d_n) = \left<\ee{\sum_k \td t_k\kappa_k}\,\,\prod_{i=1}^n \psi_i^{d_i}\right>_{g,n}.
\eeq

\subsection{Lemma}

\bl\label{mainlemma}
Let $J=\{d_1,\dots,d_n\}$, and let $2-2g-n<0$, we have
\bea
 2\,{\d W_n^{(g)}(J)\over \d B_{k,l}} 
&=& {1\over (k+1)\,(l+1)}\,\Res_{z\to \infty}\,\Res_{z'\to \infty}\, z^{k+1}\,z'^{l+1}\,\Big[W_{n+2}^{(g-1)}(z,z',J) 
\cr
&& + \sum_h\sum^{{\rm stable}}_{I\subset J} \,W_{1+\#I}^{(h)}(z,I)\,\,W_{1+n-\#I}^{(g-h)}(z',J\setminus I) \cr
&& + 2 \sum_{z_i\in J} \,W_{2}^{(0)}(z,z_i)\,\,W_{n}^{(g)}(z',J\setminus \{z_i\})\Big] \cr
\eea
which implies that
\bea
 {\d A_n^{(g)}(J)\over \d B_{2k,2l}} 
&=& 2^{-k-l-1}\,(2k-1)!!\,(2l-1)!!\,\,\Big[A_{n+2}^{(g-1)}(k,l,J) 
\cr
&& + \sum_h\sum^{{\rm stable}}_{I\subset J} \,A_{1+\#I}^{(h)}(k,I)\,\,A_{1+n-\#I}^{(g-h)}(l,J\setminus I)\Big] \, .
\eea
Moreover the derivatives of $A_n^{(g)}$ with respect to $B_{k,l}$ where $k$ or $l$ is odd vanish.

We shall denote
\beq
\hat B_{k,l} = (2k-1)!!\,(2l-1)!!\,2^{-k-l-1}\,\,B_{2k,2l}.
\eeq
\el

\proof{
We present a self contained proof of the first equation of this lemma in appendix \ref{applemma} below, but we mention that this lemma is a straightforward application of the formalism of Kostov and Orantin \cite{Kostov2010, OrantinN.2008}, based on earlier work of Kostov, and related to the Givental formalism.

It can also be seen as a very simple generalization of the "holomorphic anomaly equations as in \cite{Eynarda, EOFg}.
Let us recall that in \cite{EOFg}, modular transformations of the spectral curve amount to change the Bergman kernel as:
\beq\label{Bdukdul}
B(z_1,z_2) \to B(z_1,z_2) + \sum_{k,l} c_{k,l}\, du_k(z_1)\,du_l(z_2)
\eeq
where $du_k$, $k=1,\dots, {\rm genus}$, are the holomorphic forms on the spectral curve, satisfying:
\beq
du_k(z) = {1\over 2i\pi}\,\oint_{z'\in\bcycle_k} B(z,z'),
\eeq
and it was found in \cite{EOFg} that
\bea\label{dBdmodular}
 2\,{\d W_n^{(g)}(J)\over \d c_{k,l}} 
&=& {1\over (2i\pi)^2}\,\oint_{z\in \bcycle_k} \oint_{z'\in \bcycle_l}\, \,\Big[W_{n+2}^{(g-1)}(z,z',J) 
\cr
&& + \sum_h\sum^{{\rm stable}}_{I\subset J} \,W_{1+\#I}^{(h)}(z,I)\,\,W_{1+n-\#I}^{(g-h)}(z',J\setminus I)\Big] . \cr
\eea
It can be seen that the derivation of \cite{EOFg} doesn't rely on the fact that $du_k$ are holomorphic, it works for $du_k$ meromorphic, and thus the present Lemma is an analogous of this when $du_k(z)$ are of the form $du_k = \zeta^k\,d\zeta$.

\medskip
Therefore, similarly to \eq{Bdukdul}, we write (doing as if the sum over $k$ and $l$ were finite), and using the local parameter $z=\zeta=\sqrt{x(z)-x(a)}$:
\bea
B(z_1,z_2) -  \Brond(z_1,z_2) 
&=& \sum_{k,l} B_{k,l}\,z_1^k\,z_2^l\,\, dz_1\,dz_2  \cr
&=& \sum_{k,l}\,B_{k,l}\, \Res_{z\to\infty}\Res_{z'\to\infty}\,\, {z^{k+1}\over k+1}\,B(z,z_1)\,{z'^{l+1}\over l+1}\,B(z',z_2)  \cr
\eea
and, similarly to \eq{dBdmodular}, we get
\bea\label{eqdWngdBkl1}
 2\,{\d W_n^{(g)}(J)\over \d B_{k,l}} 
&=& {1\over (k+1)\,(l+1)}\,\Res_{z\to \infty}\,\Res_{z'\to \infty}\, z^{k+1}\,z'^{l+1}\,\Big[W_{n+2}^{(g-1)}(z,z',J) 
\cr
&& + \sum_h\sum^{'}_{I\subset J} \,W_{1+\#I}^{(h)}(z,I)\,\,W_{1+n-\#I}^{(g-h)}(z',J\setminus I)\Big] 
\eea
where $\sum'$ excludes all cases where one of the factors is $W_1^{(0)}$.

This was just a sketch of the proof, a full self contained proof of this equation is presented in appendix \ref{applemma}.

\medskip
Now, let us prove the second part of the Lemma.

Notice that the sum in the right hand side of \eq{eqdWngdBkl1} includes cases where one of the factors is $W_2^{(0)}$, i.e. we write
\bea
 2\,{\d W_n^{(g)}(J)\over \d B_{k,l}} 
&=& {1\over (k+1)\,(l+1)}\,\Res_{z\to \infty}\,\Res_{z'\to \infty}\, z^{k+1}\,z'^{l+1}\,\Big[W_{n+2}^{(g-1)}(z,z',J) 
\cr
&& + \sum_h\sum^{{\rm stable}}_{I\subset J} \,W_{1+\#I}^{(h)}(z,I)\,\,W_{1+n-\#I}^{(g-h)}(z',J\setminus I) \cr
&& + 2 \sum_{z_i\in J} \,W_{2}^{(0)}(z,z_i)\,\,W_{n}^{(g)}(z',J\setminus \{z_i\})\Big] \cr
\eea
where now $\sum^{\rm stable}$ means both factors must be stable, i.e. we exclude all terms where one factor is either $W_1^{(0)}$ or $W_2^{(0)}$.

Since $W_2^{(0)}(z,z_i)=B(z,z_i)$, the residues of the last term give
\bea
 2\,{\d W_n^{(g)}(J)\over \d B_{k,l}} 
&=& {1\over (k+1)\,(l+1)}\,\Res_{z\to \infty}\,\Res_{z'\to \infty}\, z^{k+1}\,z'^{l+1}\,\Big[W_{n+2}^{(g-1)}(z,z',J) 
\cr
&& + \sum_h\sum^{{\rm stable}}_{I\subset J} \,W_{1+\#I}^{(h)}(z,I)\,\,W_{1+n-\#I}^{(g-h)}(z',J\setminus I) \Big]\cr
&& + {2\over (l+1)}\, \sum_{z_i\in J} \,z_i^k\,dz_i\,\,\Res_{z'\to\infty} z'^{l+1}\,W_{n}^{(g)}(z',J\setminus \{z_i\}) \cr
\eea
If we write that
\beq
W_n^{(g)}(z_1,\dots,z_n) = 2^{d_{g,n}}\sum_{d_1+\dots+d_n\leq 3g-3+n}\,\, A^{(g)}_n(d_1,\dots,d_n)\,\,\prod_{i=1}^n d\xi_{d_i}(z_i)
\eeq
and using that
\beq
\Res_{z'\to \infty} z'^{2k+1}\,d\xi_d(z')=-\Res_{z'\to 0} z'^{2k+1}\,d\xi_d(z') =  (2d+1)!!\,2^{-d}\,\delta_{k,d},
\eeq
and
\beq
\Res_{z'\to \infty} z'^{2k}\,d\xi_d(z')=0,
\eeq
we find 
\bea
&& 4\sum_{d_1,\dots,d_n} \Big[ \,{\d A_n^{(g)}(d_1,\dots,d_n)\over \d B_{k,l}}\prod_i d\xi_{d_i}(z_i) + \sum_i A_n^{(g)}(d_1,\dots,d_n)\,{\d d\xi_{d_i}(z_i)\over \d B_{k,l}}\, \prod_{j\neq i} d\xi_{d_j}(z_j) \Big]\cr
&=& \sum_{d,d'}\sum_{d_1,\dots,d_n}\,\delta_{k,2d}\,\delta_{l,2d'}\,\,(2d-1)!!\,(2d'-1)!!\,2^{-d-d'}\,\Big[ A_{n+2}^{(g-1)}(d,d',d_1,\dots,d_n)
\cr
&& + \sum_h\sum^{{\rm stable}}_{I\subset \{d_1,\dots,d_n\}} \,A_{1+\#I}^{(h)}(d,I)\,\,A_{1+n-\#I}^{(g-h)}(d',{\{d_1,\dots,d_n\}\setminus I}) \Big]\,\,\prod_{i=1}^n d\xi_{d_i}(z_i)\cr
&& + 4\, \sum_{i=1}^n \,z_i^k\,dz_i\,\,\sum_{d'}\sum_{d_1,\dots,d_n} \delta_{l,2d'}\,(2d'-1)!!2^{-d'} A_{n}^{(g)}(d',\{d_1,\dots,d_n\}\setminus \{d_i\})\prod_{j\neq i} d\xi_{d_j}(z_j)  \cr
\eea
The last line exactly simplifies with the second term in the first line, and thus we get:
\bea
&& 2 \,{\d A_n^{(g)}(d_1,\dots,d_n)\over \d B_{k,l}}\prod_i d\xi_{d_i}(z_i) \cr
&=& \sum_{d,d'}\,\delta_{k,2d}\,\delta_{l,2d'}\,\,(2d-1)!!\,(2d'-1)!!\,2^{-d-d'-1}\,\Big[ A_{n+2}^{(g-1)}(d,d',d_1,\dots,d_n)
\cr
&& + \sum_h\sum^{{\rm stable}}_{I\subset \{d_1,\dots,d_n\}} \,A_{1+\#I}^{(h)}(d,I)\,\,A_{1+n-\#I}^{(g-h)}(d',{\{d_1,\dots,d_n\}\setminus I}) \Big]\,\,\cr
\eea
which is the Lemma.

}

\medskip

At $B=\Brond$, i.e. at $\hat B_{k,l}=0$, we have
\beq
\stackrel{{\circ}}{A}^{(g)}_n(d_1,\dots,d_n)= \,<\psi_1^{d_1}\dots\psi_n^{d_n}\,\,\ee{\sum_k \td t_k \kappa_k}>_{g,n}
\eeq
and thus
\bea
&& 2\,\left.{\d\over \d \hat B_{k,l}}\,{A}^{(g)}_n(J)\right|_{\hat B_{k,l}=0} 
= <\psi_{n+1}^k\psi_{n+2}^l\,\prod_{i\in J}\psi_i^{d_i}\,\,\ee{\sum_k \td t_k \kappa_k}>_{g-1,n+2} \cr
&& + \sum_h\sum^{{\rm stable}}_{I\subset J} \,<\psi_{n+1}^k\,\prod_{i\in I}\psi_i^{d_i}\,\,\ee{\sum_k \td t_k \kappa_k}>_{h,1+\#I}\,<\psi_{n+2}^l\,\prod_{i\notin I}\psi_i^{d_i}\,\,\ee{\sum_k \td t_k \kappa_k}>_{g-h,1+n-\#I}\Big] \, . \cr
&=& \left< \sum_\delta l_{\delta*} \psi_{n+1}^k\psi_{n+2}^l\,\prod_{i\in J}\psi_i^{d_i}\,\,\ee{\sum_k \td t_k \kappa_k}\right>_{g,n}.
\eea
Similarly, computing the $m^{\rm th}$ derivative at $B=\Brond$ we get:
\beq
 2^m\,\left.{\d^m\over \d \hat B_{k_1,l_1}\dots \d \hat B_{k_m,l_m}}\,{A}^{(g)}_n(J)\right|_{\hat B_{k,l}=0}  
= \left< \prod_{r=1}^m\left(\sum_\delta l_{\delta*} \psi_{n+2r-1}^{k_r}\psi_{n+2r}^{l_r}\right)\,\prod_{i\in J}\psi_i^{d_i}\,\,\ee{\sum_k \td t_k \kappa_k}\right>_{g,n}
\eeq
And thus, by writing the Taylor expansion we get
\beq
{A}^{(g)}_n(d_1,\dots,d_n)=<\psi_1^{d_1}\dots\psi_n^{d_n}\,\,\ee{\sum_k \td t_k \kappa_k}\,\, \ee{{1\over 2}l_* \hat B(\psi,\psi')}>_{g,n},
\eeq
where
\beq
\hat B(\psi,\psi') = \sum_{k,l}\,\hat B_{k,l}\,\,\psi^k\,\psi'^l,
\eeq
and $l_*=\sum_{\delta} l_{\delta*}$ is the projection to all boundary divisors $\delta$.

\medskip

This ends the proof of theorem \ref{thspinv1bp}.

\subsection{Change of basis}

It is sometimes good idea to change the basis $d\xi_d$ to another basis.
\beq
d\xi_d= \sum_{d'\leq d} C_{d,d-d'}\,d\hat\xi_{d'}.
\eeq
That gives
\bea
&& 2^{-d_{g,n}}\,W_n^{(g)}(z_1,\dots,z_n)  \cr
&=& \sum_{d_i} \prod_i d\xi_{d_i}(z_i) \,\left< \prod_i \psi_i^{d_i}\,\ee{{1\over 2}\sum_{k,l} \hat B_{k,l}\,l_* \psi^k\psi'^l}\,\ee{\sum_k \td t_k \kappa_k}\right>_{g,n} \cr
&=& \sum_{d_i,d'_i} \prod_i  C_{d_i,d_i-d'_i}\,d\hat\xi_{d'_i}(z_i) \,\left< \prod_i \psi_i^{d_i}\,\ee{{1\over 2}\sum_{k,l} \hat B_{k,l}\,l_* \psi^k\psi'^l}\,\ee{\sum_k \td t_k \kappa_k}\right>_{g,n} \cr
&=& \sum_{d'_i} \prod_i \,d\hat\xi_{d'_i}(z_i) \,\left< \prod_i \psi_i^{d'_i}\,(\sum_{d_i} C_{d_i+d'_i,d_i}\,\psi_i^{d_i})\,\ee{{1\over 2}\sum_{k,l} \hat B_{k,l}\,l_* \psi^k\psi'^l}\,\ee{\sum_k \td t_k \kappa_k}\right>_{g,n} \cr
\eea
and therefore it is interesting to introduce the functions
\beq
f_d(u) = u^{d}\, \sum_k C_{d+k} u^{k}
\eeq
that gives
\beq
W_n^{(g)}(z_1,\dots,z_n)  
=  2^{d_{g,n}}\sum_{d_i} \prod_i \,d\hat\xi_{d_i}(z_i) \,\left< \prod_i \psi_i^{d_i}\,\prod_i f_{d_i}(\psi_i)\,\ee{{1\over 2}\sum_{k,l} \hat B_{k,l}\,l_* \psi^k\psi'^l}\,\ee{\sum_k \td t_k \kappa_k}\right>_{g,n} .
\eeq
Those changes of basis are very useful for the topological vertex below.

\section{Topological vertex, proof of theorem \ref{cortopvertex}}\label{secproofvertex}

Here, we prove theorem \ref{cortopvertex} by applying theorem \ref{thspinv1bp} to the toplogical vertex.
This mostly consists in computing Laplace transforms.
\smallskip 

Consider the topological vertex curve with framing $f$ (see \cite{Aganagic2004}). The 1-leg framed topological vertex's spectral \cite{MV01} curve is $\spcurve_{{\rm vertex}}=(\mathbb C\setminus ]-\infty,0]\cup[1,+\infty[,x(z)=-f\ln z-\ln{(1-z)},y(z)=-\ln z, B(z,z')=dz\otimes dz'/(z-z')^2)$, which satisfies:
\beq
\ee{-x}=\ee{-fy}\,(1-\ee{-y}).
\eeq
It is most often written with the exponential $\mathbb C^*$ variables $X=\ee{-x}$ and $Y=\ee{-y}=z$:
\beq
X=Y^f\,(1-Y).
\eeq
The only branchpoint is at $z=a={f\over f+1}$, at which we have
\beq
X(a)=\ee{-x(a)}={f^f\over (f+1)^{f+1}}.
\eeq 

\bigskip

Just observe that changing $z\to 1/z$ is equivalent to changing $f\to -f-1$ in $x(z)$, it changes $y(z)\to -y(z)$ and it doesn't change $B(z_1,z_2)$, and therefore all $\td t_k$ and $\hat B_{k,l}$ are unchanged by changing $f\to -f-1$:
\beq\label{eqftomfm1}
\td t_k(-f-1)=\td t_k(f)
\virg
\hat B_{k,l}(-f-1) = \hat B_{k,l}(f).
\eeq

\medskip
Similarly, changing $z\to 1-z$ and $x\to 1/f x$, is equivalent to changing $f\to 1/f$. $B(z_1,z_2)$ is unchanged, but in the expansion in powers of $x-x(a)$, the change $x\to x/f$ induces powers of $f$. This changes $\td t_k\to \td t_k\,f^{2k-1}$ and $\hat B_{k,l}\to \hat B_{k,l}\,f^{k+l+1}$:
\beq\label{eqfto1overf}
\td t_k(1/f)= f^{2k-1}\,\td t_k(f)
\virg
\hat B_{k,l}(1/f) = f^{k+l+1}\,\hat B_{k,l}(f).
\eeq

Those symmetry properties are of course the consequences of the fact that $\mathbb C^3$ is a toric Calabi-Yau 3-fold.

\subsubsection{Computing $\td t_k$}

If we assume $f\in \mathbb R_+$, we have $a=f/(1+f)\in ]0,1[$, and the steepest descent contour $\gamma$ passing through the branchpoint, such that $x(\gamma)-x(a)=\mathbb R_+$, is simply
\beq
\gamma=[0,1].
\eeq
The Laplace transform $\ee{-g(u)}$ of $ydx$ is easily written in terms of the variable $z$, using \eq{eqGuinteuxdy}, and gives an Euler Beta-function:
\bea
\ee{-g(u)} 
&=&  {2\,u^{1/2}\,(f+1)^{(f+1)u}\over f^{fu}\,\sqrt\pi}\,\, \int_0^1 z^{fu}\,(1-z)^{u}\,\, dz/z \cr
&=&  {2\,u^{1/2}\,(f+1)^{(f+1)u}\over f^{fu}\,\sqrt\pi}\,\, {\Gamma(fu)\,\Gamma(1+u)\over \Gamma((f+1)u+1)}  \cr
&=&  {2\,u^{1/2}\,(f+1)^{(f+1)u}\over (f+1)\,f^{fu}\,\sqrt\pi}\,\, {\Gamma(fu)\,\Gamma(u)\over \Gamma((f+1)u)} \cr
\eea
Stirling's large $u$ expansion of the $\Gamma$ function gives (see appendix \ref{appStirling})
\beq
\ln\Gamma(u) = u\ln u - u +{1\over 2}\ln{(2\pi/u)} + \sum_{k\geq 1}\,{\Ber_{2k}\over 2k(2k-1)}\,u^{1-2k}
\eeq
where $\Ber_k$ is the $k^{\rm th}$ Bernoulli number.
That gives
\beq
\ee{\td t_0} = \sqrt{f(f+1)\over 8}
\eeq
and for $k\geq 1$, $\td t_{2k=0}$ and
\beq
\td t_{2k-1} = {\Ber_{2k}\over 2k(2k-1)}\,\, \left({1\over (f+1)^{2k-1}}-{1\over f^{2k-1}}-1\right).
\eeq
Notice that it indeed satisfies the symmetries \eq{eqftomfm1} and \eq{eqfto1overf}.


\subsubsection{Computing $\xi_d$}

We have
\beq
d\xi_0(z)=-\,\Res_{z'\to a}\, B(z,z')\,{1\over \sqrt{x(z')-x(a)}},
\eeq
or integrating once:
\beq
\xi_0(z)=\Res_{z'\to a}\, {dz'\over z-z'}\,\,{1\over \sqrt{x(z')-x(a)}}.
\eeq
The pole is a simple pole and the residue is easily computed and gives
\beq
\xi_0(z)  = {\sqrt{2\over x''(a)}}\,\,\,{1\over z-a}  = \sqrt{2f\over (1+f)^3}\,\,{1\over z-{f\over 1+f}}.
\eeq
Notice that $x'(z) = {z(1+f)-f\over z(1-z)}$, and thus we can also write
\beq\label{dxi0dzoverdx}
\xi_0(z) = \sqrt{2f\over f+1}\,\,{1\over z(1-z)}\,\,\,{dz\over dx(z)}.
\eeq

Then, for $d\geq 1$, we have
\beq\label{dxidvertex1}
\xi_d(z)=(2d-1)!!\,2^{-d}\,\Res_{z'\to a}\, {dz'\over z-z'}\,{1\over (x(z')-x(a))^{d+1/2}},
\eeq
which shows that $\xi_d(z)$ must be a rational fraction of $z$, with a pole of degree $2d+1$ at $z=a$ and no other pole, and which must behave as:
\beq
\xi_d(z) \sim {(2d-1)!!\,2^{-d}\over (x(z)-x(a))^{(d+1/2)}} + O(1) .
\eeq
Since $x'(z)$ is a rational fraction:
\beq
x'(z) = {z(1+f)-f\over z(1-z)},
\eeq
we see that $-d\xi_d(z)/dx(z)$ is also a rational fraction of $z$, and it clearly has a pole only at $z=a$, and near that pole, it behaves like (see \eq{defdxidBkl})
\beq
-d\xi_d(z)/dx(z) \sim {(2d+1)!!\,2^{-d-1}\over (x(z)-x(a))^{(d+3/2)}} + {(2d-1)!!\,2^{-d-1}\,\,B_{2d,0}\over \sqrt{x(z)-x(a)}} +  O(1) ,
\eeq
which proves that
\beq\label{recdxid}
\xi_{d+1}(z)=-\,{d\xi_d(z)\over dx(z)} - \hat B_{d,0}\,\, \xi_0(z),
\eeq
and then
\beq
\xi_d = (-1)^d\,\xi_0^{(d)} - \sum_{k=0}^{d-1} (-1)^k\,\hat B_{d-1-k,0}\,\xi_0^{(k)}=-\sum_{k=0}^{d} (-1)^k\,\hat B_{d-1-k,0}\,\xi_0^{(k)},
\eeq
where we defined $\hat B_{-1,0}=-1$, and $\xi_0^{(d)} = (d/dx)^d\,\xi_0$.
We thus have
\bea\label{Wngvertexdxideriv1}
&& 2^{-d_{g,n}}\,W_n^{(g)}(z_1,\dots,z_n)  \cr
&=& \sum_{d_i} \prod_i d\xi_{d_i}(z_i) \,\left< \prod_i \psi_i^{d_i}\,\ee{{1\over 2}\sum_{k,l} \hat B_{k,l}\,l_* \psi^k\psi'^l}\,\ee{\sum_k \td t_k \kappa_k}\right>_{g,n} \cr
&=& \sum_{d_i,d'_i} \prod_i (-1)^{d'_i} (-\hat B_{d_i-d'_i-1,0})\,d\xi_0^{(d'_i)}(z_i) \,\left< \prod_i \psi_i^{d_i}\,\ee{{1\over 2}\sum_{k,l} \hat B_{k,l}\,l_* \psi^k\psi'^l}\,\ee{\sum_k \td t_k \kappa_k}\right>_{g,n} \cr
&=& \sum_{d_i} \prod_i (-1)^{d_i}\,d\xi_0^{(d_i)}(z_i) \,\left< \prod_i( -\sum_{k\geq -1} \hat B_{k,0}\psi_i^{d_i+k+1})\,\ee{{1\over 2}\sum_{k,l} \hat B_{k,l}\,l_* \psi^k\psi'^l}\,\ee{\sum_k \td t_k \kappa_k}\right>_{g,n} \cr
&=& \sum_{d_i} \prod_i (-1)^{d_i}\,d\xi_0^{(d_i)}(z_i) \,\left< \prod_i \psi_i^{d_i} \,\, \prod_i (1-\sum_{k\geq 0} \hat B_{k,0}\psi_i^{k+1})\,\ee{{1\over 2}\sum_{k,l} \hat B_{k,l}\,l_* \psi^k\psi'^l}\,\ee{\sum_k \td t_k \kappa_k}\right>_{g,n} \cr
\eea

\subsubsection{Computation of $\hat B_{0,k}$}

writing $\zeta(z)=\sqrt{x(z)-x(a)}$, we have
\beq
d\xi_0(z) = -\,\sqrt{2\over x''(a)}\,{dz\over (z-a)^2} = -\,{d\zeta\over \zeta^2}-\sum_{k} B_{0,k}\,\zeta^k\,d\zeta.
\eeq
Let us compute the Laplace transform:
\bea
\int_\gamma \,(d\xi_0(z)+{d\zeta\over \zeta^2})\,\,\ee{-u(x(z)-x(a))} 
&=& -\,\sum_k B_{0,2k} \int_{-\infty}^{\infty} \zeta^{2k}\,d\zeta\,\ee{-u\zeta^2} \cr
&=& -\,\sum_k B_{0,2k} \,\,\sqrt\pi\,\, u^{-k-1/2}\,{(2k-1)!!\over 2^k} \cr
&=& -\,2\,\sqrt{\pi\,u}\,\,\,\sum_k \hat B_{0,k} \,\, u^{-k-1} \cr
\eea
Since $d\xi_0(z)+{d\zeta\over \zeta^2}$ is analytical at $z=a$, we may slightly deform the contour, let us say, surrounding $a$ in the upper half-plane.
We have
\bea
\int_\gamma \,{d\zeta\over \zeta^2}\,\,\ee{-u(x(z)-x(a))} 
&=& \int_{-\infty}^\infty \,{d\zeta\over \zeta^2}\,\,\ee{-u\,\zeta^2} \cr
&=& -\,\int_{-\infty}^\infty \,\,\ee{-u\,\zeta^2}\,\,d\,{1\over \zeta} \cr
&=& \,\int_{-\infty}^\infty \,{1\over \zeta}\,d\ee{-u\,\zeta^2}\,\, \cr
&=& -2u\,\int_{-\infty}^\infty \,d\zeta\,\ee{-u\,\zeta^2}\,\, \cr
&=& -2\,\sqrt{\pi\, u} \cr
\eea
and
\bea
\int_\gamma \,d\xi_0(z)\,\,\ee{-u(x(z)-x(a))} 
&=& \ee{u\,x(a)}\,\sqrt{2\over x''(a)}\,\, \int_0^1 \,\,\ee{-u\, x(z)}  \,\, d{1\over z-a} \cr
&=& u\,\,\ee{u\,x(a)}\,\sqrt{2f\over (f+1)^3}\,\, \int_0^1 \,{dx(z)\over z-a}\,\ee{-u\, x(z)}  \cr
&=& u\,\,\ee{u\,x(a)}\,\sqrt{2f\over f+1}\,\, \int_0^1 \,{dz\over z(1-z)}\,\ee{-u\, x(z)}  \cr
&=& u\,\,\ee{u\,x(a)}\,\sqrt{2f\over f+1}\,\, \int_0^1 \,{dz\over z(1-z)}\,z^{fu}\,(1-z)^u  \cr
&=& u\,\,\ee{u\,x(a)}\,\sqrt{2f\over f+1}\,\, {\Gamma(u)\,\Gamma(fu)\over \Gamma((f+1)u)}  \cr
&=& u\,\,{(f+1)^{(f+1)u}\over f^{fu}}\,\sqrt{2f\over f+1}\,\, {\Gamma(u)\,\Gamma(fu)\over \Gamma((f+1)u)}  \cr
&=& 2\,\sqrt{\pi\,u}\,\, \ee{\sum_k {\Ber_{2k}\over 2k(2k-1)}\,u^{1-2k}\,\left(1+f^{1-2k}-(f+1)^{1-2k}\right)}
\eea
Eventually, we get that
\beq
\sum_k \hat B_{0,k}\,u^{-k-1} = 1- \ee{\sum_k {\Ber_{2k}\over 2k(2k-1)}\,u^{1-2k}\,\left(1+f^{1-2k}-(f+1)^{1-2k}\right)} = 1-\ee{-\sum_{k> 0}\td t_k u^{-k}}=1-\ee{-g(u)}.
\eeq
where we have redefined $g(u)$ without the term $\td t_0$.

According to \eq{Wngvertexdxideriv1}, we thus have:
\bea
&& 2^{-d_{g,n}}\,\ee{\td t_0\chi_{g,n}}\,W_n^{(g)}(z_1,\dots,z_n)  \cr
&=& \sum_{d_i} \prod_i (-1)^{d_i}\,d\xi_0^{(d_i)}(z_i) \,\left< \prod_i \psi_i^{d_i} \,\, \,\ee{{1\over 2}\sum_{k,l} \hat B_{k,l}\,l_* \psi^k\psi'^l}\,\ee{\sum_{k>0} \td t_k (\kappa_k-\sum_{i=1}^n \psi_i^k)}\right>_{g,n} .\cr
\eea


\subsubsection{Computation of $\hat B_{k,l}$}

Following \eq{defdxidBkl}, we write
\beq
\xi_0(z) = {1\over \zeta} - \sum_{k} B_{0,k} {\zeta^{k+1}\over k+1}
\eeq
and thus
\bea
\xi_0^{(j)}(z) 
&=& \left({d\over 2\zeta\,d\zeta}\right)^{j}\,\xi_0(z) \cr
&=& {(-1)^j\,(2j-1)!!\over 2^j\,\zeta^{2j+1}} 
- \sum_{k} B_{0,k} \,{(k-1)(k-3)\dots(k-1-2j)\over 2^j}\,\zeta^{k-2j+1}\cr
\eea
and that implies by the recursion \eq{recdxid}
\bea
\xi_{d}(z) 
&=& {(2d-1)!!\,2^{-d}\over \zeta^{2d+1}} \cr
&& - (-1)^d\,\sum_{l} B_{0,l+2d}\, \zeta^{l+1}\,\,(l+2d-1)(l+2d-3)\dots(l+3)\,2^{-d} \cr
&& -\sum_{k=0}^{d-1} \sum_l (-1)^k\,\hat B_{d-1-k,0} B_{0,l}  \zeta^{l-2k+1}\,\,(l-1)\dots(l-2k+3)\,2^{-k} \cr
\eea
and comparing with \eq{defdxidBkl}
\beq
\xi_{d}(z) 
= {(2d-1)!!\,2^{-d}\over (x(z)-x(a))^{d+{1\over 2}}} - (2d-1)!!\,2^{-d}\sum_{l} B_{2d,l} {(x(z)-x(a))^{l+1\over 2}\over l+1}
\eeq
we get:
\bea
(2d-1)!!\,B_{2d,l} 
&=& (-1)^d\,B_{0,l+2d}\,\,(l+2d-1)(l+2d-3)\dots(l+1)  \cr
&& + \sum_{k=0}^{d-1} (-1)^k\,2^{d-k}\,\hat B_{d-1-k,0}\,B_{0,l+2k}\,\,(l+2k-1)(l+2k-3)\dots(l+1)  \cr
\eea
and therefore
\beq
\hat B_{d,l}
= (-1)^d\,\hat B_{0,d+l}  + \sum_{k=0}^{d-1} (-1)^k\,\hat B_{d-1-k,0}\,\hat B_{0,l+k}
\eeq
Let us define the generating functions
\beq
\sum_{k\geq -1} \hat B_{0,k} u^{-k-1} = 1-\sum_{k\geq 0} \hat B_{0,k} u^{-k-1} = \ee{-g(u)}
\eeq
and we remind that we have found that $g(-u)=-g(u)$.
We have
\bea
\sum_{k\geq 0}\,\sum_{l\geq 0} \hat B_{k,l}\, u^{-k}\,v^{-l} 
&=& \sum_{k\geq 0}\,\sum_{l\geq 0}\, \sum_{j=0}^k\, (-1)^j\,\hat B_{0,k-j-1}\,\hat B_{0,l+j}\, u^{-k}\,v^{-l} \cr
&=& \sum_{l\geq 0}\, \sum_{j\geq 0}\,\sum_{k\geq j}\, (-1)^j\,\hat B_{0,k-j-1}\,\hat B_{0,l+j}\, u^{-k}\,v^{-l} \cr
&=& \sum_{l\geq 0}\, \sum_{j\geq 0}\,\sum_{k\geq -1}\, (-1)^j\,\hat B_{0,k}\,\hat B_{0,l+j}\, u^{-k-1-j}\,v^{-l} \cr
&=& -\, \ee{-g(u)}\,\sum_{l\geq 0}\, \sum_{j\geq 0}\,\, (-1)^j\,\,\hat B_{0,l+j}\, u^{-j}\,v^{-l} \cr
&=& -\, \ee{-g(u)}\,\sum_{m\geq 0}\, \hat B_{0,m}\,v^{-m}\,\sum_{j= 0}^m\,\, (-1)^j\,\, u^{-j}\,v^{j} \cr
&=& -\, \ee{-g(u)}\,u\,v\,\sum_{m\geq 0}\, \hat B_{0,m}\,\,{v^{-m-1}+(-1)^{m}\,u^{-m-1}\over u+v} \cr
&=&  \ee{-g(u)}\,u\,v\,\,{\ee{-g(v)}-\ee{-g(-u)}\over u+v} .
\eea
Finally, the generating function of $\hat B_{k,l}$ is:
\beq
\sum_{k\geq 0}\,\sum_{l\geq 0} \hat B_{k,l}\, u^{-k}\,v^{-l} 
=  u\,v\,\,{\ee{-g(u)}\,\ee{-g(v)}-1\over u+v} .
\eeq
This shows that the Laplace transform \eq{eqdefBhatuv} of $B$ is
\beq
\hat B(u,v) = {1\over 2}\,\,{\ee{-g(u)}\,\ee{-g(v)}\over 1/u+1/v}
\eeq

\subsubsection{Rewriting using intersection numbers identities}

Now, let us rewrite $\sum_{k,l} \hat B_{k,l} \psi^k\,\psi'^l$ using lemma \ref{lemmarelintnumbers} in appendix \ref{applemmarelintnumbers}.
At each step we have to compute
\beq
\sum_{k,l} \hat B_{k,l} \left< \psi^k\,\psi'^l \,\ee{\sum_k \td t_k\kappa_k}\,\, \Psi\right>_{g,n+2}
\eeq
where $\Psi$ is some polynomial in $\psi_1,\dots, \psi_n$, in particular $\Psi$ doesn't involve any $\kappa$ class.
We write
\bea
&& \sum_{k,l} \hat B_{k,l}\, \psi^{k}\,\psi'^{l} \cr
&=& -\,\ee{-g(1/\psi)}\,\, {\ee{g(1/\psi)}-\ee{g(-1/\psi')}\over \psi+\psi'}  \cr
&=& -\,\ee{-g(1/\psi)}\,\, \sum_m \sum_{j_1,\dots,j_m} {\td t_{2j_1+1}\dots \td t_{2j_m+1}\over m!}\,\,\sum_{k=0}^{m-1+2\sum j_i} (-1)^k\,\psi^{m-1-k+2\sum j_i}\,\psi'^k \cr
\eea
The first identity of Lemma \ref{lemmarelintnumbers} allows to replace it by
\beq
 -\,\, \sum_m \sum_{j_1,\dots,j_m} {\td t_{2j_1+1}\dots \td t_{2j_m+1}\over m!}\,\,\sum_{k=0}^{m-1+2\sum j_i} (-1)^k\,\kappa_{m-2-k+2\sum j_i} \,\psi'^k
\eeq
The derivative with respect to $\td t_{2j+1}$ is
\beq
 -\,\, \sum_m \sum_{j_1,\dots,j_{m-1}} {\td t_{2j_1+1}\dots \td t_{2j_{m-1}+1}\over (m-1)!}\,\,\sum_{k=0}^{m+2j+2\sum j_i} (-1)^k\,\kappa_{m-2-k+2j+2\sum j_i} \,\psi'^k
\eeq
and the second identity of Lemma \ref{lemmarelintnumbers} allows to replace it by
\beq
 -\,\,\sum_{k=0}^{2j} (-1)^k\,\psi^{2j-k} \,\psi'^k
\eeq
I.e. we have
\beq
\sum_{k,l} \hat B_{k,l} \left< \psi^k\,\psi'^l \,\ee{\sum_k \td t_k\kappa_k}\,\, \Psi\right>_{g,n+2}
= \sum_j \sum_k \td t_{2j+1}\,\,(-1)^k\,\left< \psi^{2j-k} \,\psi'^k  \,\ee{\sum_k \td t_k\kappa_k}\,\, \Psi\right>_{g,n+2}
\eeq

\subsubsection{The Hodge class}

We thus see that the spectral curve's class appearing in theorem \ref{thspinv1bp}, is the product of 3 classes:
\beq
\left<\ee{\sum_{k>0} \td t_k( \kappa_k-\sum_i \psi_i^k)}\,\ee{{1\over 2}\sum_\delta \sum_{k,l} \hat B_{k,l}\,l_{\delta*}\,\psi^k\,\psi'^l}\right>_{g,n}
= \left<\Lambda(1)\Lambda(f)\Lambda(-1-f)\right>_{g,n}
\eeq
where
\beq
\Lambda(f) = \ee{-\sum_k {1\over f^{2k-1}}\,\,{B_{2k}\over 2k(2k-1)}\,\left(\kappa_{2k-1}-\sum_{i=1}^n \psi_i^{2k-1} + {1\over 2}\,\sum_\delta \sum_{l=0}^{2k-2}\,(-1)^l \,l_{\delta*}\,\psi^l\,\psi'^{2k-2-l}\right)}
\eeq

Using Mumford's formula \cite{Mumford1983}, we recognize the Hodge class.
\beq
\Lambda(f) = \sum_k (-1)^k\,f^{-k} c_{k}(\mathbb E) = {\rm Hodge\, class}.
\eeq

Theorem \ref{thspinv1bp}, then says that, for the topological vertex's spectral curve, we have
\bea
&& W_n^{(g)}(z_1,\dots,z_n)  \cr
&=& 2^{d_{g,n}}\,\ee{-\td t_0\,\chi_{g,n}}\sum_{d_1,\dots,d_n} \prod_i (-1)^{d_i}d\xi_0^{(d_i)}(z_i)\,\, \left< \psi_1^{d_1}\dots\psi_n^{d_n}\, \Lambda(1)\Lambda(f)\Lambda(-f-1)\right>_{g,n} .\cr
\eea
In other words, we have re--proved that the "remodelling the B-model" proposal of Bouchard-Klemm-Mari\~no-Pasquetti (BKMP conjecture \cite{Mar1,BKMP}) is valid for the topological vertex.
This theorem was in fact already proved by Chen \cite{ChenLin2009} and Zhou \cite{ZhouJian2009}, using cut and join equations.

\subsubsection{Laplace transform and Mari\~no--Vafa form}

Let write $\xi_0(z)$ in Laplace transform
\beq
\xi_{0}(z) = \sum_{\mu=0}^\infty C_\mu\,\, \ee{-\mu\,x(z)}= \sum_{\mu=0}^\infty C_\mu\,\, X(z)^{\mu},
\eeq
This is equivalent to a Taylor expansion near $z=1$, in powers of $X(z)=\ee{-x(z)}=z^f\,(1-z)$.
We thus have
\bea
C_\mu
&=& \Res_{z\to 1}\,\, \xi_{0}(z)\,\,X(z)^{-\mu}\,{dX(z)\over X(z)} \cr
&=& {\sqrt{2}\over \sqrt{f(f+1)}}\, \Res_{z\to 1}\,{1\over (f+1)z-f}\, X(z)^{-\mu}\,\left( {-f\,dz\over z}+{dz\over 1-z}\right) \cr
&=& {\sqrt{2}\over \sqrt{f(f+1)}}\, \Res_{z\to 1}\,{1\over (f+1)z-f}\, X(z)^{-\mu}\,{( z-f(1-z))dz\over z(1-z)} \cr
&=& {\sqrt{2}\over \sqrt{f(f+1)}}\, \Res_{z\to 1}\,\, X(z)^{-\mu}\,{dz\over z(1-z)} \cr
&=& {\sqrt{2}\over \sqrt{f(f+1)}}\, \Res_{z\to 1}\,\, {1\over z^{\mu f}\,(1-z)^{\mu}}\,{dz\over z(1-z)} \cr
&=& -\,{\sqrt{2}\over \sqrt{f(f+1)}}\, {\Gamma(1+\mu(f+1))\over \mu!\,\Gamma(1+\mu f)} \cr
&=& -\,{\sqrt{2}\,(f+1)\over f\,\sqrt{f(f+1)}}\, {\Gamma(\mu(f+1))\over \mu!\,\Gamma(\mu f)}
\eea
This implies
\beq
\xi_{0}(z) = -\,
{\sqrt{2}\,(f+1)\over f\,\sqrt{f(f+1)}}\, \sum_\mu \ee{-\mu x(z)}\,\,{\Gamma(\mu(f+1))\over \mu!\,\Gamma(\mu f)}
\eeq
and taking derivatives:
\beq
d\xi_0^{(d)}(z) = -\,(-\mu)^{d+1}\,dx(z)\,\,{\sqrt{2}\,(f+1)\over f\,\sqrt{f(f+1)}}\, \sum_\mu \ee{-\mu x(z)}\,\,{\Gamma(\mu(f+1))\over \mu!\,\Gamma(\mu f)}.
\eeq
Then, write
\beq
\sum_{d_i} (-\mu_i)^{d_i+1}\,\psi_i^{d_i} = {-\mu_i\over 1+\mu_i\psi_i}
\eeq
That gives the Laplace transform of $W_n^{(g)}$ as:
\bea
W_n^{(g)}(z_1,\dots,z_n) 
&=& \,2^{d_{g,n}}\,\ee{-\td t_0\chi_{g,n}}\,(2(f+1)/f^3)^{n/2}\,\sum_{\mu_1,\dots,\mu_n} \cr
&& \prod_{i=1}^n {\Gamma(\mu_i(f+1))\,\over \mu_i!\,\Gamma(\mu_i f)}\,\, \mu_i\,\ee{-\mu_i x(z_i)}\,dx(z_i)\cr
&& \left< \prod_{i=1}^n{1\over 1+\mu_i\psi_i}\, \,\,\Lambda(1)\Lambda(f)\Lambda(-f-1)\right>_{g,n}
\eea
which is the famous Mari\~no--Vafa formula \cite{MV01, Liu2003a}.

\section{Examples}

Let us show a few more examples.

\subsection{Example: Weil-Petersson}

Choose the Weil-Petersson curve:
\beq
y={1\over 2\pi}\,\sin(2\pi\sqrt x)
\eeq
or more precisely ${\cal S}_{\rm W.P}=(\mathbb C, x(z)=z^2,y(z)={1\over 2\pi}\sin{(2\pi z)},B=\Brond)$.
It has only one branchpoint at $z=a=0$, and $B(z,z')=\Brond(z,z')=dz\otimes dz'/(z-z')^2$.

We have
\bea
G(u)=\ee{-g(u)} 
&=&  {2u^{1/2}\,\over \sqrt\pi}\,\, \int\, \ee{-u x}\,\, dy  \cr
&=&  {2u^{1/2}\,\over  \sqrt\pi}\,\, \int_{-\infty}^\infty\, \ee{-u z^2}\,\, \cos{(2\pi z)}\, dz \cr
&=& \,\, {u^{1/2}\,\over  \sqrt\pi}\,\, \int_{-\infty}^\infty\, \ee{-u z^2}\,\, (\ee{2i\pi z}+\ee{-2i\pi z})\, dz \cr
&=&  2{u^{1/2}\,\over  \sqrt\pi}\,\, \int_{-\infty}^\infty\, \ee{-u z^2}\,\, \ee{2i\pi z}\, dz \cr
&=&  2{u^{1/2}\,\over  \sqrt\pi}\,\, \int_{-\infty}^\infty\, \ee{-u (z-i\pi/u)^2}\,\, \ee{-\pi^2/u}\, dz \cr
&=&   2\,\ee{-\pi^2/u} \cr
\eea
i.e.
\beq
g(u) = -\ln 2+ \pi^2/u
\eeq
and thus
\beq
\ee{\sum_k \td t_k \kappa_k} = 2^{-\kappa_0}\, \ee{\pi^2\,\kappa_1}.
\eeq
We also have
\beq
\xi_d(z) = {(2d-1)!!\over 2^d\,\,z^{2d+1}},
\eeq
and thus
\bea
&& W_n^{(g)}(z_1,\dots,z_n) \cr
&=& (-1)^n\,2^{d_{g,n}+\chi_{g,n}}\,\sum_{d_i} \prod_{i=1}^n {(2d_i+1)!!\,dz_i\over 2^{d_i}\,z_i^{2d_i+2}}\,\,\left<\ee{\pi^2\kappa_1}\prod_i \psi_i^{d_i}\right>_{g,n} \cr
&=& (-1)^n\,2^{d_{g,n}+\chi_{g,n}}\!\!\!\!\!\sum_{d_0+d_1+\dots+d_n=d_{g,n}} \prod_{i=1}^n {(2d_i+1)!!\,dz_i\over 2^{d_i}\,z_i^{2d_i+2}}\,\,\left<{(\pi^2\kappa_1)^{d_0}\over d_0!}\prod_{i=1}^n \psi_i^{d_i}\right>_{g,n} \cr
&=& (-1)^n\,2^{\chi_{g,n}}\!\!\!\!\!\sum_{d_0+d_1+\dots+d_n=d_{g,n}} \prod_{i=1}^n {(2d_i+1)!!\,dz_i\over \,z_i^{2d_i+2}}\,\,\left<{(2\pi^2\kappa_1)^{d_0}\over d_0!}\prod_{i=1}^n \psi_i^{d_i}\right>_{g,n} \cr
&=& (-1)^n\,2^{\chi_{g,n}}\!\!\!\!\!\sum_{d_0+d_1+\dots+d_n=d_{g,n}} \prod_{i=1}^n {(2d_i+1)!\,dz_i\over 2^{d_i}\,d_i!\,z_i^{2d_i+2}}\,\,\left<{(2\pi^2\kappa_1)^{d_0}\over d_0!}\prod_{i=1}^n \psi_i^{d_i}\right>_{g,n} \cr
\eea
Notice that
\beq
\int_0^\infty L\,dL\,\,L^{2d}\,\ee{-zL} = {(2d+1)!\over z^{2d+2}}
\eeq
therefore
\bea
&& {W_n^{(g)}(z_1,\dots,z_n)\over dz_1\dots dz_n} \cr
&=& (-1)^n\,2^{\chi_{g,n}}\!\!\!\!\!\sum_{d_0+d_1+\dots+d_n=d_{g,n}} \prod_{i=1}^N\int_0^\infty L_i dL_i \ee{-z_i L_i}\,\,\prod_{i=1}^n {L_i^{2d_i}\over 2^{d_i}\,d_i!}\,\,\left<{(2\pi^2\kappa_1)^{d_0}\over d_0!}\prod_{i=1}^n \psi_i^{d_i}\right>_{g,n} \cr
&=& {(-1)^n\,2^{\chi_{g,n}}\over d_{g,n}!} \prod_{i=1}^N\int_0^\infty L_i dL_i \ee{-z_i L_i}\,\,\,\,\left<(2\pi^2\kappa_1+{1\over 2}\sum_{i=1}^n L_i^2\psi_i)^{d_{g,n}}\right>_{g,n} \cr
&=& (-1)^n\,2^{\chi_{g,n}} \prod_{i=1}^N\int_0^\infty L_i dL_i \ee{-z_i L_i}\,\,\,\,\left<\ee{2\pi^2\kappa_1+{1\over 2}\sum_{i=1}^n L_i^2\psi_i}\right>_{g,n} \cr
\eea
It is known (see Wolpert \cite{Wolpert1983}) that $2\pi^2\kappa_1$ is the Weil-Petersson form. In the Fenchel-Nielsen coordinates $(l_i,\theta_i)$ in Teichm\"uller space:
\beq
2\pi^2 \kappa_1 = \sum_i dl_i\wedge d\theta_i,
\eeq
and thus, we have rederived that the symplectic invariants are the Laplace transform of the Weil Petersson volumes
\beq
{\rm Vol}(L_1,\dots,L_n) = \left<\ee{2\pi^2\kappa_1+{1\over 2}\sum_{i=1}^n L_i^2\psi_i}\right>_{g,n}.
\eeq
The fact that symplectic invariants satisfy the topological recursion, is equivalent \cite{Mulase2006, Eynard2007, Liu2007} (after Laplace transform), to the fact that Weil-Petersson volumes satisfy Mirzakhani's recursion relation \cite{Mirzakhani2006}.

\subsection{Example: Lambert curve}\label{secHurwitz}

Choose the Lambert curve $(\mathbb C\setminus \mathbb R_-, x(z)=-z+\ln z, y(z)=z, B=dz_1\otimes dz_2/(z_1-z_2)^2)$, i.e. $y$ as a function of $\ee{x}$ is the Lambert function:
\beq
\ee{x}=y\ee{-y} \qquad \leftrightarrow \qquad y=L(\ee{x}).
\eeq
We have
\beq
dx=(-1+{1\over z})\,dz,
\eeq
and thus there is a unique branchpoint (solution of $dx=0$) at $a=1, y=1, x=-1$. 

\medskip

In principle, all the computations about the Lambert curve can be obtained by taking the $f\to\infty$ limit in the topological vertex \cite{BM, MV01, Liu2003a}, however, for completeness, let us rederive it directely.

\subsubsection{The times $\td t_k$}

\begin{figure}[h]
\centering
\includegraphics[width=6cm]{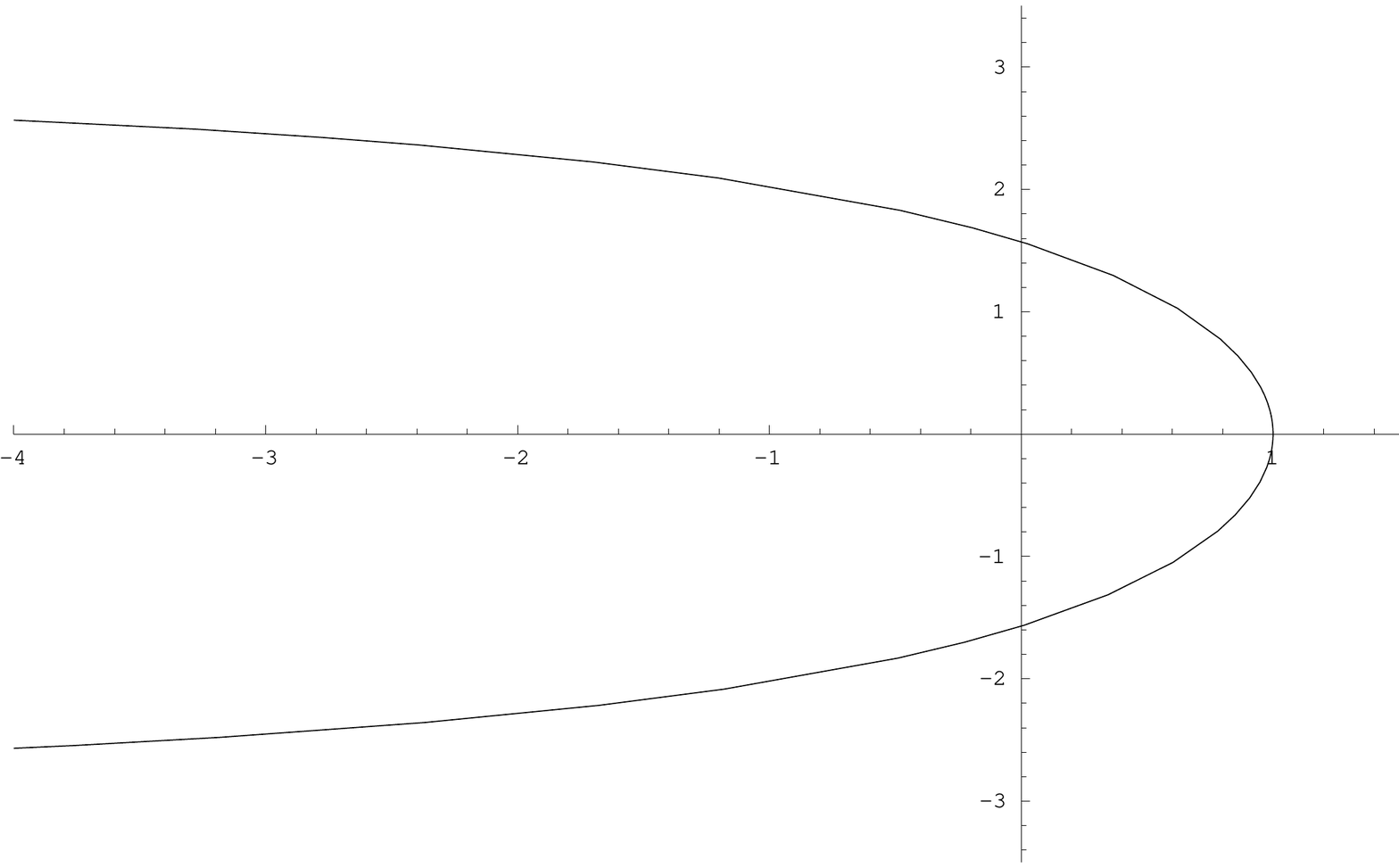}
\caption{\footnotesize{The steepest descent path for the Lambert curve. It surrounds the negative real axis. In polar coordinates, it has equation $\rho=\theta/\sin\theta$.}}
\label{figgammalambert}
\end{figure}

The steepest descent path $\gamma$ such that $x(\gamma)=[-1,+\infty[$, can be written in polar coordinates $z=\rho\ee{i\theta}$, as $\rho=\theta/\sin\theta$, see fig. \ref{figgammalambert}. It is easy to see that $\gamma$ can be deformed into a contour surrounding the negative real axis $\mathbb R_-$. 

We have:
\bea
\ee{-g(u)} 
&=&  {2\,u^{1/2}\,\ee{-u}\over \sqrt\pi}\,\, \int_{\gamma} (y\,\ee{-y})^{-u}\,\,  dy \cr
&=&  {2\,u^{1/2}\,\ee{-u}\over \sqrt\pi}\,\, \int_{\gamma} y^{-u}\,\ee{uy}\,\,  dy \cr
&=& 4i\sin{\pi u}\,\, {u^{1/2}\,\ee{-u}\over \sqrt\pi}\,\,  \int_{0}^\infty y^{-u}\,\ee{-uy}\,\,  dy \cr
&=& 4i\sin{\pi u}\,\, {u^{1/2}\,\ee{-u}\over \sqrt\pi}\,\, u^{u-1}\, \Gamma(1-u) \cr
&=& 4i\,\sqrt\pi\,\, u^{-1/2}\,\ee{-u}\,\, u^{u}\, {1\over \Gamma(u)} \cr
\eea
From the Stirling expansion:
\beq
\ln\Gamma(u) = u\ln u - u +{1\over 2}\,\ln(2\pi/u) + \sum_{k=1}^\infty {\Ber_{2k}\over 2k(2k-1)}\,u^{1-2k}
\eeq
we find
\beq
\td t_0=-{1\over 2}\,\ln 8 + {i\pi\over 2}
\virg
\td t_{2k-1} =  {\Ber_{2k}\over 2k(2k-1)}.
\eeq
We thus have to consider:
\beq
\ee{\sum_{k\geq 1} \td t_k\kappa_k}= \ee{-\sum_{k=1}^\infty {\Ber_{2k}\over 2k(2k-1)}\,\kappa_{2k-1}} = 1 - {\kappa_1\over 12} + \dots
 \eeq

\subsubsection{Computing $\xi_d$}

Like in section \ref{secproofvertex}, we have
\beq
\xi_0(z)=\Res_{z'\to a}\, {dz'\over z-z'}\,\,{1\over \sqrt{x(z')-x(a)}}
 = {\sqrt{2\over x''(a)}}\,\,\,{1\over z-a}  = {-i\,\sqrt{2}\over z-1}.
\eeq
Notice that $x'(z) = {1-z\over z}$, and thus we can also write
\beq
\xi_0(z) = {i\,\sqrt{2}\over z}\,\,\,{dz\over dx(z)}.
\eeq
And like in section \ref{secproofvertex}, 
since $x'(z)={1-z\over z}$ is a rational fraction with a zero only at $z=a=1$,
we see that $-d\xi_d(z)/dx(z)$ is also a rational fraction of $z$, and it clearly has a pole only at $z=a$, and near that pole, it behaves like (see \eq{defdxidBkl})
\beq
-d\xi_d(z)/dx(z) \sim {(2d+1)!!\,2^{-d-1}\over (x(z)-x(a))^{(d+3/2)}} + {(2d-1)!!\,2^{-d-1}\,\,B_{2d,0}\over \sqrt{x(z)-x(a)}} +  O(1) ,
\eeq
which proves that
\beq
\xi_{d+1}(z)=-\,{d\xi_d(z)\over dx(z)} - \hat B_{d,0}\,\, \xi_0(z),
\eeq
and then
\beq
\xi_d = (-1)^d\,\xi_0^{(d)} - \sum_{k=0}^{d-1} (-1)^k\,\hat B_{d-1-k,0}\,\xi_0^{(k)}=-\sum_{k=0}^{d} (-1)^k\,\hat B_{d-1-k,0}\,\xi_0^{(k)},
\eeq
where we defined $\hat B_{-1,0}=-1$, and $\xi_0^{(d)} = (d/dx)^d\,\xi_0$.
We thus have, like in section \ref{secproofvertex}
\bea
&& 2^{-d_{g,n}}\,W_n^{(g)}(z_1,\dots,z_n)  \cr
&=& \sum_{d_i} \prod_i (-1)^{d_i}\,d\xi_0^{(d_i)}(z_i) \,\left< \prod_i \psi_i^{d_i} \,\, \prod_i (1-\sum_{k\geq 0} \hat B_{k,0}\psi_i^{k+1})\,\ee{{1\over 2}\sum_{k,l} \hat B_{k,l}\,l_* \psi^k\psi'^l}\,\ee{\sum_k \td t_k \kappa_k}\right>_{g,n} \cr
\eea

\subsubsection{Computation of $\hat B_{0,k}$}

Like in section \ref{secproofvertex}, we have
\beq
\int_\gamma \,(d\xi_0(z)+{d\zeta\over \zeta^2})\,\,\ee{-u(x(z)-x(a))} 
= -\,2\,\sqrt{\pi\,u}\,\,\,\sum_k \hat B_{0,k} \,\, u^{-k-1} .
\eeq
Since $d\xi_0(z)+{d\zeta\over \zeta^2}$ is analytical at $z=a$, we may slightly deform the contour, let us say, surrounding $a$ in the upper half-plane.
We have
\beq
\int_\gamma \,{d\zeta\over \zeta^2}\,\,\ee{-u(x(z)-x(a))} 
= -2\,\sqrt{\pi\, u} 
\eeq
and
\bea
\int_\gamma \,d\xi_0(z)\,\,\ee{-u(x(z)-x(a))} 
&=& \ee{u\,x(a)}\,\sqrt{2\over x''(a)}\,\, \int_\gamma \,\,\ee{-u\, x(z)}  \,\, d{1\over z-1} \cr
&=& u\,\,\ee{-u}\,i\,\sqrt{2}\,\, \int_\gamma \,{dx(z)\over z-1}\,\ee{-u\, x(z)}  \cr
&=& -\,u\,i\sqrt{2}\,\ee{-u}\,\,\, \int_\gamma \,{dz\over z}\,\ee{-u\, x(z)}  \cr
&=& -\,u\,i\sqrt{2}\,\ee{-u}\,\,\, \int_\gamma \,{dz\over z}\,\ee{u\, z}\,z^{-u}  \cr
&=& 2\sin{(\pi u)}\,u\,i\sqrt{2}\,\ee{-u}\,\,\, \int_0^\infty \,{dz\over z}\,\ee{-u\, z}\,z^{-u}  \cr
&=& 2\sin{(\pi u)}\,u\,i\sqrt{2}\,\ee{-u}\,\,\, u^{u}\,\Gamma(-u) \cr
&=& 2\pi\,u\,i\sqrt{2}\,\ee{-u}\,\,\, u^{u}\,{1\over \Gamma(1+u)} \cr
\eea
Eventually, we get that
\beq
\sum_k \hat B_{0,k}\,u^{-k-1} = 1- \ee{\sum_k {\Ber_{2k}\over 2k(2k-1)}\,u^{1-2k}\,}=1-\ee{-g(u)}.
\eeq
where we have redefined $g(u)$ without the term $\td t_0$.

We thus have:
\bea
&& 2^{-d_{g,n}}\,\ee{\td t_0\chi_{g,n}}\,W_n^{(g)}(z_1,\dots,z_n)  \cr
&=& \sum_{d_i} \prod_i (-1)^{d_i}\,d\xi_0^{(d_i)}(z_i) \,\left< \prod_i \psi_i^{d_i} \,\, \,\ee{{1\over 2}\sum_{k,l} \hat B_{k,l}\,l_* \psi^k\psi'^l}\,\ee{\sum_{k>0} \td t_k (\kappa_k-\sum_i \psi_i^k)}\right>_{g,n} .\cr
\eea

Then, all the same steps as in section \ref{secproofvertex} give that 
 the generating function of $\hat B_{k,l}$ is:
\beq
\sum_{k\geq 0}\,\sum_{l\geq 0} \hat B_{k,l}\, u^{-k}\,v^{-l} 
=  u\,v\,\,{\ee{-g(u)}\,\ee{-g(v)}-1\over u+v} .
\eeq

And, using lemma \ref{lemmarelintnumbers} as in section \ref{secproofvertex}, we get

\beq
\sum_{k,l} \hat B_{k,l} \left< \psi^k\,\psi'^l \,\ee{\sum_k \td t_k\kappa_k}\,\, \Psi\right>_{g,n+2}
= \sum_j \sum_k \td t_{2j+1}\,\,(-1)^k\,\left< \psi^{2j-k} \,\psi'^k  \,\ee{\sum_k \td t_k\kappa_k}\,\, \Psi\right>_{g,n+2}.
\eeq

\subsubsection{The Hodge class}

We thus see that the spectral curve's class appearing in theorem \ref{thspinv1bp}, is:
\beq
\left<\ee{\sum_{k>0} \td t_k( \kappa_k-\sum_i \psi_i^k)}\,\ee{{1\over 2}\sum_\delta \sum_{k,l} \hat B_{k,l}\,l_{\delta*}\,\psi^k\,\psi'^l}\right>_{g,n}
= \left<\Lambda(1)\right>_{g,n}
\eeq
where
\beq
\Lambda(1) = \ee{-\sum_k \,\,{B_{2k}\over 2k(2k-1)}\,\left(\kappa_{2k-1}-\sum_{i=1}^n \psi_i^{2k-1} + {1\over 2}\,\sum_\delta \sum_{l=0}^{2k-2}\,(-1)^l \,l_{\delta*}\,\psi^l\,\psi'^{2k-2-l}\right)}
\eeq
i.e.
using Mumford's formula \cite{Mumford1983}, we recognize the Hodge class.
\beq
\Lambda(f) = \sum_k (-1)^k\,f^{-k} c_{k}(\mathbb E) = {\rm Hodge\, class}.
\eeq

Theorem \ref{thspinv1bp}, then says that, for the Lambert spectral curve, we have
\bea
&& W_n^{(g)}(z_1,\dots,z_n)  \cr
&=& 2^{d_{g,n}}\,\ee{-\td t_0\,\chi_{g,n}}\sum_{d_1,\dots,d_n} \prod_i (-1)^{d_i}d\xi_0^{(d_i)}(z_i)\,\, \left< \psi_1^{d_1}\dots\psi_n^{d_n}\, \Lambda(1)\right>_{g,n} .\cr
\eea
In other words, we have re--proved the Bouchard-Mari\~no conjecture \cite{BM}.
This theorem was in fact already proved in \cite{Borot1} using a matrix model, and in \cite{Eynardb} using cut and join equations.

\subsubsection{Laplace transform and ELSV form}

Let us Laplace transform $\xi_0(z)$, i.e. expand it near $z=0$, in powers of $X(z)=\ee{x(z)}=z\,\ee{-z}$:
\beq
\xi_{0}(z) = \sum_{\mu=0}^\infty C_\mu\,\, \ee{\mu\,x(z)}= \sum_{\mu=0}^\infty C_\mu\,\, X(z)^{\mu}.
\eeq

We have
\bea
C_\mu
&=& \Res_{z\to 0}\,\, \xi_{0}(z)\,\,X(z)^{-\mu}\,{dX(z)\over X(z)} \cr
&=& {i\sqrt{2}}\, \Res_{z\to 0}\,{1\over z}\,{dz\over dx(z)}\, X(z)^{-\mu}\,dx(z) \cr
&=& {i\sqrt{2}}\, \Res_{z\to 0}\,{dz\over z}\,\, X(z)^{-\mu}\, \cr
&=& {i\sqrt{2}}\, \Res_{z\to 0}\,{dz\over z}\,\, z^{-\mu}\,\, \ee{\mu z}\, \cr
&=& {i\sqrt{2}}\, {\mu^\mu\over \mu!} \cr
\eea
This implies
\beq
\xi_{0}(z) = i\,\sqrt{2}\,\, \sum_\mu \ee{\mu x(z)}\,\,{\mu^\mu\over \mu!}.
\eeq
and taking derivatives:
\beq
d\xi_0^{(d)}(z) 
= i\,\sqrt{2}\,\, \sum_\mu \ee{\mu x(z)}\,\,{\mu^\mu\over \mu!}\,\mu^{d+1}\,\, dx(z)\, .
\eeq
Then, write
\beq
\sum_{d_i} \mu_i^{d_i+1}\,\psi_i^{d_i} = {\mu_i\over 1-\mu_i\psi_i}
\eeq
That gives the Laplace transform of $W_n^{(g)}$ as:
\bea
W_n^{(g)}(z_1,\dots,z_n) 
&=& \,2^{d_{g,n}}\,\ee{-\td t_0\chi_{g,n}}\,(-2)^{n/2}\,\sum_{\mu_1,\dots,\mu_n} 
 \prod_{i=1}^n {\mu_i^{\mu_i}\,\over \mu_i!}\,\, \mu_i\,\ee{\mu_i x(z_i)}\,dx(z_i)\cr
&& \left< \prod_{i=1}^n{1\over 1-\mu_i\psi_i}\, \,\,\Lambda(1)\right>_{g,n},
\eea
which is the famous ELSV formula \cite{EKEDAHL1999, Ekedahl2001} for Hurwitz numbers.

\subsection{Matrix models and Hankel class}

Formal matrix model are generating functions enumerating discrete surfaces.
Their correlation functions are defined as power series in $t$:
\bea
\om_n^{(g)}(x_1,\dots,x_n;t;t_3,\dots,t_d)
= \sum_{v=1}^\infty t^v \sum_{S\in \mathbb M_{g,n}(v)}\,{1\over \#{\rm Aut}(S)}\,\,{t_3^{n_3(S)}t_4^{n_4(S)}\dots t_d^{n_d(S)}\over x_1^{1+l_1(S)}\dots x_n^{1+l_n(S)}}
\eea
where $\mathbb M_{g,n}(v)$ is the finite set of oriented discrete surfaces (also called "maps", see \cite{Berge2003, W.T.Tutte1963}), made of polygonal faces of degree between $3$ and $d$, of genus $g$, and with $v$ vertices, and with $n$ polygonal marked faces (and each marked face having one oriented marked edge).
If $S\in \mathbb M_{g,n}(v)$, we call $n_j(S)$ the number of unmarked faces of degree $j$ (and we have $j\geq 3$), we call $l_i(S)$ the degree of the $i^{\rm th}$ marked face, and $\#{\rm Aut}(S)$ the cardinal of the automorphism group of $S$.

Most often, the dependence on $t;t_3,\dots,t_d$ will be implicitely understood, and we write
\beq
\om_n^{(g)}(x_1,\dots,x_n)\equiv \om_n^{(g)}(x_1,\dots,x_n;t;t_3,\dots,t_d).
\eeq

It was proved in \cite{Eynard2004} that the generating functions $W_n^{(g)}=\om_n^{(g)}dx_1\dots dx_n$ satisfy the topological recursion, with a spectral curve given by \cite{W.T.Tutte1963}:
\beq
\spcurve_{\rm Matrix}=
\left\{\begin{array}{l}
{\cal C}=\mathbb C^* \cr
x(z) = \alpha+\gamma(z+1/z) \cr
y(z) = \sum_{k=1}^{d-1}\, u_k z^{-k} \cr
B(z_1,z_2) = {dz_1\otimes dz_2\over (z_1-z_2)^2}
\end{array}\right.
\eeq
where the coefficients $\alpha, \gamma$ and $u_k$ are determined by:
\beq
\left\{\begin{array}{l}
\sum_k u_k (z^k+z^{-k}) = x(z)-\sum_{j=3}^d t_j\, x(z)^{j-1} \cr
u_0=0 \cr
u_1 = {t\over \gamma}
\end{array}\right.
\eeq
and we choose the unique solution such that $\gamma^2 = t+O(t^2)$ and $\alpha=O(t)$.

\medskip
$\bullet$ {\bf Example Quadrangulations}

we choose $t_4\neq 0$ and all other $t_j=0$, that gives
\beq
\left\{\begin{array}{l}
\gamma^2 = {1-\sqrt{1-12 t t_4}\over 6  t_4}\virg \alpha=0 \cr
u_1={t\over \gamma} \virg u_2=0\virg u_3 = - t_4 \gamma^3
\end{array}\right.
\eeq
and thus
\beq
\spcurve_{\rm Quadrangulations}=
\left\{\begin{array}{l}
{\cal C}=\mathbb C^* \cr
x(z) = \gamma(z+1/z) \cr
y(z) = {t\over \gamma z} -  t_4 \gamma^3\,z^{-3}\cr
B(z_1,z_2) = {dz_1\otimes dz_2\over (z_1-z_2)^2}
\end{array}\right.
\eeq

\medskip

Solving $x'(z)=0$, we see that
those spectral curves have 2 branchpoints, located at $z=a=\pm 1$.
the case of multiple branchpoints will be done in a coming paper, but for the moment, let us compute the spectral curve class associated to the branch point at $a=1$.

\smallskip
Assuming $\alpha$ and $\gamma$ real positive, The steepest descent path going through $a=1$, is simply $\gamma=[0,\infty[$.
The Laplace transform gives
\bea
\int_\gamma \ee{-u\,(x(z)-x(a))}\,dy(z)
&=& \gamma\int_0^\infty \ee{-u\gamma(z+1/z-2)}\,\,\sum_k k\,u_k z^{-k}\,\,{dz\over z} \cr
&=& \gamma\,\ee{2u\gamma}\,\sum_k k\,u_k\,\,\int_{-\infty}^\infty \ee{-2u\gamma\,\cosh\phi}\,\, \ee{-k\phi}\,\,d\phi \cr
&=& \pi\,\gamma\,\,\ee{2u\gamma}\,\sum_k k\,u_k\,\,H_k(2i\,\gamma\,\,u) 
\eea
where $H_k$ is the $k^{\rm th}$ Hankel function of the 1st kind (which is closely related to the Bessel function).
Therefore
\beq
\ee{-\sum_k \td t_k u^{-k}}  = 2\,\gamma\,\,\sqrt{\pi\,u}\,\,\,\ee{2u\gamma}\,\sum_k k\,u_k\,H_k(2i\,\gamma\,\,u)
\eeq

We also have
\beq
\xi_0(z) = \sqrt{2\over x''(a)}\,{1\over z-a} = {1\over \sqrt\gamma}\,\,{1\over z-1}
\eeq
and
\beq
\xi_d(z) = {(2d-1)!!\over 2^d\,\,\gamma^{d+1/2}}\,\sum_{k=0}^{2d}\, {1\over (z-1)^{2d+1-k}}\,\,\left(\begin{array}{c}d+1/2\cr k\end{array}\right).
\eeq


This computation can be in principle pursued, and would give the number of quadrangulations (or other discrete surfaces) in terms of intersection numbers. This will be the purpose of another work.

\subsubsection{Example: resolved conifold}

On can also try to apply the general formula to the Resolved connifold's spectral curve, in order to check the BKMP conjecture.

The conifold's spectral curve $\spcurve$ is $\spcurve=(\mathbb C,x(z)=-f\ln z+\ln{(1-z)}-\ln{(1-qz)}, y(z)=-\ln z, B(z_1,z_2)=dz_1dz_2/(z_1-z_2)^2)$, it satisfies
\beq
\ee{-x}=\ee{-fy}\,{1-\ee{-y}\over 1-q\,\ee{-y}}.
\eeq
It is most often written with the exponential variables $X=\ee{-x}$ and $Y=\ee{-y}=z$, as:
\beq
X=Y^f\,{1-Y\over 1-q\,Y}.
\eeq
There are 2 branchpoints, $a_+>0$ and $a_-<\ln q$. We assume $0<q<1$, and thus the steepest descent paths for the Laplace transforms are $z\in\gamma_+=[0,1]$, and $z\in\gamma_-=]1/q,\infty[$.

The Laplace transforms $\ee{-g_\pm(u)}$ of $ydx$ are easily written in terms of the variable $z$, and give hypergeometric functions of $q$:
\bea
\ee{-g_\pm(u)} 
&=&  {2u^{1/2}\,\ee{u\,a_\pm }\over \,\sqrt\pi}\,\, \int_{\gamma_\pm} z^{fu}\,(1-z)^{u}(1-q\,z)^{-u}\,\, dz/z \cr
\eea

Thus
\bea
\ee{-g_+(u)} 
&=&  {2u^{1/2}\,\ee{u\,a_+ }\over \,\sqrt\pi}\,\, \int_0^1 z^{fu}\,(1-z)^{u}(1-q\,z)^{-u}\,\, dz/z \cr
&=&  {2u^{1/2}\,\ee{u\,a_+ }\over \,\sqrt\pi}\,\,{\Gamma(fu)\Gamma(u+1)\over \Gamma(fu+u+1)}\, {}_2F_1(u,fu;fu+u+1;q) \cr
\eea
and, by a simple change of variable $z\to q/z$:
\bea
\ee{-g_-(u)} 
&=&  {2u^{1/2}\,\ee{u\,a_- }\over \,\sqrt\pi}\,\, \int_{1/q}^\infty z^{fu}\,(1-z)^{u}(1-q\,z)^{-u}\,\, dz/z \cr
&=& \ee{-g_+(-u)}.
\eea

However, it is not so simple to compute explicitly the large $u$ expansion of $g_\pm(u)$, and this computation will be pursued in other works.

\section{Conclusion}

We have found the interpretation of symplectic invariants of a spectral curve, in terms of integrals over the moduli-space ${\cal M}_{g,n}$.

With this formula, we have found new proofs of the Bouchard-Mari\~no conjecture \cite{BM} for Hurwitz numbers, and BKMP conjecture \cite{BKMP} for $\mathbb C^3$.
We hope that the extension of the formula for several branchpoints, could help prove the BKMP conjecture for more complicated toric geometries, but there is still some substantial work ahead.

\medskip
{\bf Remarks about Mirror symmetry}

$\bullet$ Intersection numbers "count" complex curves with marked points, in some moduli-space of curves.
They are related to a type A topological string theory.
The moduli which appear in the intersection numbers are the $\td t_k$ and $\hat B_{k,l}$'s and $d\xi_d(z)$. 

\smallskip

$\bullet$ On the other hand,
symplectic invariants are defined in terms of moduli of the spectral curve, and in particular in terms of the Bergman kernel $B(z_1,z_2)$ and in terms of the 1-form $ydx$.
They are obtained by computing residues, i.e. in terms of the complex geometry on the spectral curve.
They can be thought of as a type B topological string theory.

\smallskip
We see that the relationship between the type A moduli and the type B moduli, is the Laplace transform, for instance:
\beq
\ee{-\sum_k \td t_k u^{-k}} = {2\,u^{3/2}\,\ee{ux(a)}\over \sqrt\pi}\,\int_{\gamma_a}\,\ee{-ux}\,ydx
\eeq
relates the moduli $\td t_k$ of $\kappa$--classes to the 1-form $ydx$.
The moduli of $\psi$ classes, encoded in $d\xi_d$ and in $\hat B_{k,l}$, are related to the Laplace transform of the Bergman kernel.

\smallskip
Notice also that the steepest descent contour $\gamma_a$, defined as
\beq
\Im\, x(\gamma) = {\rm constant}
\eeq
or equivalently
\beq
{\rm Arg}\, (X(\gamma))={\rm Arg}\, (\ee{-x(\gamma)}) = {\rm constant}
\eeq
is closely related to the definition of Lagrangian submanifolds. Indeed, write
\beq
X = |X|\,\ee{-\theta},
\eeq
the steepest descent contour $\gamma_a$ is a contour along which $d\theta=0$.

\bigskip

We thus see that there seems to be a deep link  between this computation, and mirror symmetry, but this link is still to be clarified.

Namely, it seems important to understand how the Laplace transform of $ydx$ is related to $\kappa$--classes, and the Laplace transform of $B$ is related to $\psi$--classes !

\medskip

\section*{Acknowledgments}
I would like to thank G. Borot, A. Brini, M. Mari\~no, M. Mulase, N. Orantin, B. Safnuk
for useful and fruitful discussions on this subject.
This work  is partly supported by the Enigma European network MRT-CT-2004-5652, by the ANR project G\'eom\'etrie et int\'egrabilit\'e en physique math\'ematique ANR-05-BLAN-0029-01,
by the European Science Foundation through the Misgam program,
by the Quebec government with the FQRNT, and the CERN for its support and hospitality.


\setcounter{section}{0}
\appendix{}
\setcounter{section}{0}

\section{Some relationships among intersection numbers}\label{applemmarelintnumbers}

\bl\label{lemmarelintnumbers}
We have the following identities for intersection numbers

\beq
 \,\left<\kappa_{d}\,\,\,\prod_{i=1}^n \psi_i^{d_i}\right>_{g,n}  
= \left<\psi_{n+1}^{d+1}\,\ee{-\sum_j \td t_j \psi_{n+1}^j}\,\,\ee{\sum_k \td t_k \kappa_k}\,\prod_{i=1}^n \psi_i^{d_i}\right>_{g,n+1}  ,
\eeq

\beq
\left<\psi_{n+1}^{d+1}\,\,\ee{\sum_k \td t_k \kappa_k}\,\prod_{i=1}^n \psi_i^{d_i}\right>_{g,n+1}  
= \sum_{m}\sum_{j_1,\dots,j_m} {\td t_{j_1}\dots \td t_{j_m}\over m!}\,\left<\kappa_{d+\sum j_i}\,\,\ee{\sum_k \td t_k \kappa_k}\,\prod_{i=1}^n \psi_i^{d_i}\right>_{g,n}  .
\eeq

\el

\proof{
Those identities can be deduced from direct geometric properties of tautological classes, similar to \cite{Arbarello1996}.

However, let us show a proof based on general properties of symplectic invariants specialized to theorem \ref{thspinvKonts}.

\medskip

Let us consider an infinitesimal variation of spectral curve:
\beq
y\to y+\delta y
\eeq
in other words, since $y(z)=\sum_k t_{k+2}z^k$:
\beq
t_k\to t_k+\delta t_k.
\eeq
This induces a variation of the times $\td t_k$ through Laplace transform:
\beq
\delta(\ee{-g(u)}) = -\delta g(u)\,\,\,\ee{-g(u)} = {2\,u^{3/2}\over \sqrt\pi}\,\, \int \ee{-ux}\,\delta y\,\,\,dx.
\eeq

\medskip
$\bullet$ Let us consider a function $\delta y(z) = -\,y^{(-d)}(z)$ such that:
\beq
\left({d\over d x(z)}\right)^{d}\,\delta y(z) = -\,y(z) \virg x(z)=z^2,
\eeq
i.e. more explicitly
\beq\label{eqdeltay1}
y^{(-d)}(z) = \,\sum_k t_{k+2}\,{\,2^d\over(k+2)(k+4)\dots (k+2d)}\,\,z^{k+2d}.
\eeq

By integration by parts we compute $\delta g$:
\beq
\delta g(u)\,\,\,\ee{-g(u)} = u^{-d}\,\,{2\,u^{3/2}\over \sqrt\pi}\,\, \int \ee{-ux}\, y\,\,\,dx = u^{-d}\,\,\ee{-g(u)}
\eeq
i.e.
\beq
\delta g(u) = u^{-d},
\eeq
i.e.
\beq
\td t_k \to\,\,\td t_k+ \delta_{k,d}\,\delta \td t_d,
\eeq
i.e. our infinitesimal variation $\delta$ is in fact
\beq
\delta = {\d\over \d \td t_d}.
\eeq

On the other hand, we compute the dual cycle to the variation $\delta y$ of the spectral curve as (form--cycle duality is realized by the Bergman kernel):
\beq
\delta y(z)\,\,dx(z) = \,\Res_{z'\to \infty}\, B(z,z')\, y^{(-d-1)}(z')
\eeq
and the special geometry property of symplectic invariants then implies that:
\beq
\delta W_n^{(g)}(z_1,\dots,z_n) = \,\Res_{z'\to \infty}\, W_{n+1}^{(g)}(z_1,\dots,z_n,z')\, y^{(-d-1)}(z'),
\eeq
and since the only poles of $W_{n+1}^{(g)}$ are at the branchpoint $z=0$, we may move the integration contour and get:
\beq
\delta W_n^{(g)}(z_1,\dots,z_n) = -\,\Res_{z'\to 0}\, W_{n+1}^{(g)}(z_1,\dots,z_n,z')\, y^{(-d-1)}(z').
\eeq

From theorem \ref{thspinvKonts}, we thus have
\bea
&& (-1)^n\,2^{3g-3+n}\,{\d\over \d \td t_d}\,\left<\ee{\sum_k \td t_k \kappa_k}\,\,\prod_{i=1}^n \psi_i^{d_i}\right>_{g,n} \cr
&=&-\,(-1)^{n+1}\,2^{3g-3+n+1}\,\Res_{z'\to 0} y^{(-d-1)}(z')\,\,\sum_{d'} {(2d'+1)!!\,dz'\over 2^{d'}\,z'^{2d'+2}}\,\left<\psi_{n+1}^{d'}\,\ee{\sum_k \td t_k \kappa_k}\,\,\prod_{i=1}^n \psi_i^{d_i}\right>_{g,n+1}\cr
\eea
using \eq{eqdeltay1}, the residues give:
\bea
 \left<\kappa_d\,\ee{\sum_k \td t_k \kappa_k}\,\,\prod_{i=1}^n \psi_i^{d_i}\right>_{g,n} 
&=& \sum_{k} {t_{2k+3}\, (2k+1)!!\,\over 2^{k}}\,\left<\psi_{n+1}^{k+d+1}\,\ee{\sum_k \td t_k \kappa_k}\,\,\prod_{i=1}^n \psi_i^{d_i}\right>_{g,n+1} \cr
&=& \left<\psi_{n+1}^{d+1}\,\ee{-\sum_k \td t_k \psi_{n+1}^k}\,\ee{\sum_k \td t_k \kappa_k}\,\,\prod_{i=1}^n \psi_i^{d_i}\right>_{g,n+1}.
\eea
This ends the proof of the first identity.
Notice that when all $\td t_k=0$, this identity is well known \cite{Arbarello1996}.

\bigskip

$\bullet$ Now let us prove the other identity. We choose
\beq
\delta y(z) = z^{2d+1}.
\eeq
That gives
\beq
-\delta g(u)\,\ee{-g(u)} = {2u^{3/2}\over \sqrt\pi}\,\int \ee{-ux}\,\,x^{d+1/2}\,\,dx = {(2d+1)!!\over 2^{d-1}\,\,u^d}
\eeq
i.e.
\bea
\delta g(u) 
&=& -\,{(2d+1)!!\over 2^{d-1}\,\,u^d}\,\ee{g(u)} \cr
&=& -\,{(2d+1)!!\over 2^{d-1}}\,\,\sum_m \,\sum_{j_1,\dots,j_m}\, {\td t_{j_1}\dots \td t_{j_m}\over m!}\,\,u^{\sum j_i-d}  \cr
\eea
which implies
\beq
\delta \ee{\sum_k \td t_k \kappa_k} = -\,{(2d+1)!!\over 2^{d-1}}\,\,\sum_m \,\sum_{j_1,\dots,j_m}\, {\td t_{j_1}\dots \td t_{j_m}\over m!}\,\,\kappa_{\sum j_i-d}\,\ee{\sum_k \td t_k \kappa_k},
\eeq
and thus
\bea
&& \delta\,W_n^{(g)}(z_1,\dots,z_n) \cr
&=& -\,(-1)^n\,2^{3g-3+n}\,{(2d+1)!!\over 2^{d-1}}\,\,\sum_m \,\sum_{j_1,\dots,j_m}\, {\td t_{j_1}\dots \td t_{j_m}\over m!}\,\, \sum_{d_1,\dots,d_n} \cr
 && \prod_{i=1}^n {(2d_i+1)!!\,dz_i\over 2^{d_i}\,z_i^{2d_i+2}}\,\left< \kappa_{\sum j_i-d}\,\ee{\sum_k \td t_k \kappa_k}\,\,\prod_i \psi_i^{d_i} \right>_{g,n}.\cr
\eea

On the other hand, we can compute $\delta W_n^{(g)}$ from the special geometry property. 
The dual of $\delta y$ is given by:
\beq
\delta y(z)\,\,dx(z) = {-2\over 2d+3}\,\Res_{z'\to \infty} B(z,z')\,\,\, z'^{2d+3}
\eeq
and thus
\bea
&&\delta W_n^{(g)}(z_1,\dots,z_n) \cr
&=& {-2\over 2d+3}\,\Res_{z'\to \infty}\, W_{n+1}^{(g)}(z_1,\dots,z_n,z')\,  z'^{2d+3} \cr
&=& \,(-1)^{n+1}\,2^{3g-3+n+1}\,{2\,(2d+3)!!\over (2d+3)\,2^{d+1}}
 \sum_{d_1,\dots,d_n} \prod_{i=1}^n {(2d_i+1)!!\,dz_i\over 2^{d_i}\,z_i^{2d_i+2}}\,\left< \,\ee{\sum_k \td t_k \kappa_k}\,\,\prod_{i=1}^n \psi_i^{d_i} \,\,\psi_{n+1}^{d+1}\right>_{g,n+1}.\cr
\eea
Comparing those two expressions of $\delta W_n^{(g)}$ completes the proof.

}

\section{Table of intersection numbers}\label{appkappaclasses}

We organize them by Euler characteristics.

\begin{tabular}{|r||l|l|l|l|l|l|}
\hline
$-\chi $& $\,\, d_{g,n}\,\to\,\,\,$0 & 1 & 2 & 3 & 4 & 5  \\
\hline \hline
1& $<1>_{0,3}=1$ & $<\tau_1>_{1,1}={1\over 24}$ &  &  &  &   \\
  &  & $<\kappa_1>_{1,1}={1\over 24}$ &  &  &  &   \\
\hline
2  &  & $<\tau_1>_{0,4}=1$ &  $<\tau_2>_{1,2} ={1\over 24}$ &  $<\kappa_3>_{2,0}={1\over 3^2\,2^6}$&  &   \\
  &  & $<\kappa_1>_{0,4}=1$ & $<\tau_1^2>_{1,2}={1\over 24}$  &  $<\kappa_2\kappa_1>_{2,0}={1\over 5\,\,3\,\,2^3}$  &  &   \\
  &  &  & $<\tau_1\kappa_1>_{1,2}={1\over 12}$  & $<\kappa_1^3>_{2,0}={43\over 5\,\,3^2\,2^5}$  &  &   \\
  &  &  & $<\kappa_2>_{1,2}={1\over 24}$  &  &  &   \\
  &  &  & $<\kappa_1^2>_{1,2}={1\over 8}$  &  &  &   \\
  \hline
3  &  & &  $<\tau_2>_{0,5}=1$&  $<\tau_3>_{1,3}={1\over 24}$ & $<\tau_4>_{2,1}={1\over 3^2\,\,2^7}$ &     \\
  &  &  & $<\tau_1^2>_{0,5}=2$& $<\tau_2\tau_1>_{1,3}={1\over 12}$   & $<\tau_3\kappa_1>_{2,1}={29\over 5\,\,3^2\,\,2^7}$  &     \\
  &  & & $<\tau_1\kappa_1>_{0,5}=3$& $<\tau_1^3>_{1,3}={1\over 12}$   & $<\tau_2\kappa_2>_{2,1}={29\over 5\,\,3^2\,\,2^7}$  &     \\
  &  & & $<\kappa_2>_{0,5}=1$& $<\tau_2\kappa_1>_{1,3}={1\over 6}$   & $<\tau_2\kappa_1^2>_{2,1}={139\over 5\,\,3^2\,\,2^7}$  &     \\
  &  & & $<\kappa_1^2>_{0,5}=5$& $<\tau_1^2\kappa_1>_{1,3}={1\over 4}$   &  $<\tau_1\kappa_3>_{2,1}={1\over 3\,\,2^7}$  &     \\
  &  &  & & $<\tau_1\kappa_2>_{1,3}={1\over 8}$   &  $<\tau_1\kappa_2\kappa_1>_{2,1}={101\over 5\,\,3^2\,\,2^7}$  &     \\
  &  &  & & $<\tau_1\kappa_1^2>_{1,3}={13\over 24}$   &  $<\tau_1\kappa_1^3>_{2,1}={169\over 5\,\,3\,\,2^7}$  &     \\
  &  &  & & $<\kappa_3>_{1,3}={1\over 24}$   & $<\kappa_4>_{2,1}={1\over 3^2\,\,2^7}$   &     \\
  &  &  & & $<\kappa_2 \kappa_1>_{1,3}={1\over 4}$   &  $<\kappa_3\kappa_1>_{2,1}={39\over 5\,\,3^2\,\,2^7}$  &     \\
  &  &  & & $<\kappa_1^3>_{1,3}={7\over 36}$   &   $<\kappa_2^2>_{2,1}={53\over 5\,\,3^2\,\,2^7}$ &     \\
  &  &  &   & & $<\kappa_2\kappa_1^2>_{2,1}={777\over 5\,\,3^3\,\,2^7}$   &     \\
  &  &  &   & & $<\kappa_1^4>_{2,1}={29\over 2^7}$   &     \\
\hline
\end{tabular}

\medskip
A few easy  general relations are
\beq
<\tau_1^{n-3}>_{0,n}=(n-3)!
\eeq
\beq
<\tau_1^{k}\,\, \Psi>_{0,n}={(n-3)!\over (n-3-k)!}\,<\Psi>_{n-k}
\eeq

\beq
<\tau_1^{n-5}\,\, \tau_2>_{0,n}={(n-3)!\over 2}
\virg
<\tau_1^{n-6}\,\, \tau_3>_{0,n}={(n-3)!\over 3!}
\eeq
\beq
<\tau_1^{n-7}\,\, \tau_2^2>_{0,n}={(n-3)!\over 4!}\,6
\eeq

\beq
<\tau_{3g-2}>_{g,1}=<\kappa_{3g-3}>_{g,0} = {1\over 24^g\,g!}
\eeq

\section{Stirling approximation}\label{appStirling}

We have
\beq
\Gamma(u) = \int_0^\infty dz\,z^{u-1}\,\ee{-z}\,dz
\eeq
And it has the large $u$ asymptotic expansion
\beq
\ln\Gamma(u) = u\ln u - u + {1\over 2}\ln{(2\pi/u)} + \sum_{k=1}^\infty {B_{2k}\over 2k(2k-1)}\,\,{1\over u^{2k-1}}
\eeq
where $B_k$ is the $k^{\rm th}$ Bernoulli number:
\beq
B_2={1\over 6}\, , \,\, B_4={-1\over 30}\, , \,\, B_6={1\over 42}\, , \,\, B_8={-1\over 30}\, , \,\, B_{10}={5\over 66}\, , \,\, B_{12}={-691\over 2730}\,,\,\dots
\eeq

\bigskip

The Euler Beta function is:
\beq
B(u,v) = \int_0^1 dz\,z^{u-1}\,(1-z)^{v-1} = { \Gamma(u)\Gamma(v)\over \Gamma(u+v)}.
\eeq

\section{Proof of Lemma \ref{mainlemma}}\label{applemma}

We prove it by recursion on $2g-2+n$.

We shall always use the local parameter $z=\zeta=\sqrt{x(z)-x(a)}$. We have, in the small $z$ expansion:
\beq
B(z_1,z_2) ={dz_1\otimes dz_2\over (z_1-z_2)^2}+ \sum_{k,l} B_{k,l}\,z_1^k\,z_2^l\,\, dz_1\otimes dz_2  .
\eeq

First, notice that the recursive definition of $W_n^{(g)}$ involves computing $2g-2+n$ residues each containing a Bergman kernel, and also some residues may involve one or two $W_2^{(0)}=B$.
Eventually, we see that $W_n^{(g)}$ is a polynomial in the $B_{k,l}$'s of degree at most $d_{g,n}=3g-3+n$, and also, since we compute residues at each step, Taylor series near $z=0$ can be truncated to the order of poles, and this means that each $W_n^{(g)}$ involves only a finite number of $B_{k,l}$'s.

Therefore, there is no loss of generality in assuming that only a finite number of $B_{k,l}$'s are non-vanishing.
Let us also assume for the moment that $B_{k,l}$ and $B_{l,k}$ are independent variables, but in the end we will have to choose $B_{k,l}=B_{l,k}$.

\medskip

Our goal is to prove by recursion that:
\bea
\left( {\d \over \d B_{l,k}}+
\,{\d \over \d B_{k,l}}\right)\,W_{n}^{(g)}(J)  
&=&  \Res_{z\to\infty}\Res_{z'\to\infty}\,\, {z^{k+1}\over k+1}\,\,{z'^{l+1}\over l+1}\
\,\Big[ W_{n+2}^{(g-1)}(z,z',J) \cr
&& + \sum_h\sum'_{I\subset J}\, W_{1+\#I}^{(h)}(z,I)\,W_{1+n-\#I}^{(g-h)}(z',J\setminus I)\Big] .
\eea

\subsubsection*{Initialization of the recursion}

Notice that
\beq
-\,\Res_{z'\to\infty} B(z,z') {z'^{k+1}\over k+1} = z^k\,dz
\eeq
Therefore we have
\bea
{\d B(z_1,z_2) \over \d B_{k,l}} 
&=& z_1^k\,dz_1\otimes z_2^l\,dz_2 \cr
&=&  \Res_{z\to\infty}\Res_{z'\to\infty}\,\, {z^{k+1}\over k+1}\,B(z,z_1)\,{z'^{l+1}\over l+1}\,B(z',z_2)  .
\eea
This is the initial case $2g-2+n=0$ for the recursion:
\bea
&& 
\left({\d\over \d B_{k,l}}+{\d\over \d B_{l,k}}\right)\, W_2^{(0)}(z_1,z_2) \cr
&=&  \Res_{z\to\infty}\Res_{z'\to\infty}\,\, {z^{k+1}\over k+1}\,\,{z'^{l+1}\over l+1}\,\Big[W_2^{(0)}(z,z_1)W_2^{(0)}(z',z_2) +W_2^{(0)}(z,z_2)W_2^{(0)}(z',z_1)\Big] . \cr
\eea

%

This implies for the recursion kernel $K(z_0,z)$ defined in \eq{defK}:
\beq
{\d K(z_0,z_1) \over \d B_{k,l}} =  \Res_{z\to\infty}\Res_{z'\to\infty}\,\, {z^{k+1}\over k+1}\,\,{z'^{l+1}\over l+1}\,\, B(z_0,z)\,K(z',z_1).
\eeq

Assume that we have proved the lemma for every $2g'-2+n'<2g-2+n$.
We have (where $J=\{z_1,\dots,z_n\}$):
\beq
W_{n+1}^{(g)}(z_0,J) = \Res_{z''\to 0} K(z_0,z'')\,\Big[ W_{n+2}^{(g-1)}(z'',-z'',J)
+ \sum_h\sum'_{I\subset J}\, W_{1+\#I}^{(h)}(z'',I)\,W_{1+n-\#I}^{(g-h)}(-z'',J\setminus I)\Big]
\eeq
and thus
\bea
&& {\d \over \d B_{k,l}}\,W_{n+1}^{(g)}(z_0,J)  \cr
&=& \Res_{z''\to 0} {\d \over \d B_{k,l}}\,K(z_0,z'')\,\Big[ W_{n+2}^{(g-1)}(z'',-z'',J) \cr
&& + \sum_h\sum'_{I\subset J}\, W_{1+\#I}^{(h)}(z'',I)\,W_{1+n-\#I}^{(g-h)}(-z'',J\setminus I)\Big] \cr
&& + \Res_{z''\to 0} K(z_0,z'')\,\Big[ {\d \over \d B_{k,l}}\,W_{n+2}^{(g-1)}(z'',-z'',J) \cr
&& + \sum_h\sum'_{I\subset J}\, {\d \over \d B_{k,l}}\,W_{1+\#I}^{(h)}(z'',I)\,W_{1+n-\#I}^{(g-h)}(-z'',J\setminus I)\cr
&& +W_{1+\#I}^{(h)}(z'',I)\,{\d \over \d B_{k,l}}\,W_{1+n-\#I}^{(g-h)}(-z'',J\setminus I)\Big] \cr
\eea
The first term, with ${\d }\,K(z_0,z'')/ \d B_{k,l}$ gives simply
\beq
 \Res_{z\to\infty}\Res_{z'\to\infty} {z^{k+1}\over k+1}\,\,{z'^{l+1}\over l+1}\,\,B(z_0,z)\,W_{n+1}^{(g)}(z',J)
\eeq
Let us now focus on the second term, i.e. $\d/\d B_{k,l}+\d/\d B_{l,k}$ of the bracket. From the recursion hypothesis, it gives
\bea
%
(1)&|&  \Res_{z\to\infty}\Res_{z'\to\infty} {z^{k+1}\over k+1}\,\,{z'^{l+1}\over l+1}\,\,\Big[ W_{n+4}^{(g-2)}(z'',-z'',z,z',J)  \qquad \qquad \qquad \qquad \qquad  \hfill (1)\cr
(2)&|& + \sum_{h',I'\subset J} W_{3+\# I'}^{(h')}(z'',-z'',z,I')\,W_{1+n-\#I'}^{(g-1-h')}(z',J\setminus I') \cr
(3)&|& + \sum_{h',I'\subset J} W_{2+\# I'}^{(h')}(z'',z,I')\,W_{2+n-\#I'}^{(g-1-h')}(-z'',z',J\setminus I') \cr
(4)&|& + \sum_{h',I'\subset J} W_{2+\# I'}^{(h')}(-z'',z,I')\,W_{2+n-\#I'}^{(g-1-h')}(z'',z',J\setminus I') \cr
(5)&|& + \sum_{h',I'\subset J} W_{1+\# I'}^{(h')}(z,I')\,W_{3+n-\#I'}^{(g-1-h')}(z'',-z'',z',J\setminus I') \cr
(6)&|& + \sum_{h,I\subset J} W_{3+\# I}^{(h-1)}(z'',z,z',I)\,W_{1+n-\#I}^{(g-h)}(-z'',J\setminus I) \cr
(7)&|& + \sum_{h,I\subset J}\sum_{h',I'\subset I} W_{2+\# I'}^{(h')}(z'',z,I')\,W_{1+\#I-\#I'}^{(h-h')}(z',I\setminus I')\,W_{1+n-\#I}^{(g-h)}(-z'',J\setminus I) \cr
(8)&|& + \sum_{h,I\subset J}\sum_{h',I'\subset I} W_{1+\# I'}^{(h')}(z,I')\,W_{2+\#I-\#I'}^{(h-h')}(z'',z',I\setminus I')\,W_{1+n-\#I}^{(g-h)}(-z'',J\setminus I) \cr
(9)&|& + \sum_{h,I\subset J} W_{1+\# I}^{(h)}(z'',I)\,W_{3+n-\#I}^{(g-h-1)}(z,z',-z'',J\setminus I) \cr
(10)&|& + \sum_{h,I\subset J}\sum_{h',I'\subset J\setminus I} W_{1+\# I}^{(h)}(z'',I)\,W_{2+\#I'}^{(h')}(z,-z'',I')\,W_{1+n-\#I-\#I'}^{(g-h-h')}(z',(J\setminus I)\setminus I') \cr
(11)&|& + \sum_{h,I\subset J}\sum_{h',I'\subset J\setminus I} W_{1+\# I}^{(h)}(z'',I)\,W_{1+\#I'}^{(h')}(z,I')\,W_{2+n-\#I-\#I'}^{(g-h-h')}(-z'',z',(J\setminus I)\setminus I') \Big]\cr
\eea
Now, we multiply by $K(z_0,z'')$ and take the residue at $z''\to 0$, then, by definition of $W_n^{(g)}$'s, terms $(2)+(7)+(10)$ give
\beq
\sum_{h',I'\subset J} W_{2+\#I'}^{(h')}(z_0,z,I')\,W_{1+n-\#I'}^{(g-h')}(z',J\setminus I'),
\eeq
 terms $(5)+(8)+(11)$ give
\beq
\sum_{h',I'\subset J} W_{1+\#I'}^{(h')}(z,I')\,W_{2+n-\#I'}^{(g-h')}(z_0,z',J\setminus I'),
\eeq
and terms $(1)+(3)+(4)+(6)+(9)$ give
\beq
W_{3+n}^{(g-1)}(z_0,z,z',J).
\eeq
And thus finally:
\bea
&& \left({\d \over \d B_{k,l}}+{\d \over \d B_{l,k}}\right)\,W_{n+1}^{(g)}(z_0,J)  \cr
&=&  \Res_{z\to\infty}\Res_{z'\to\infty} {z^{k+1}\over k+1}\,\,{z'^{l+1}\over l+1}\,\,\Big[ 
W_{3+n}^{(g-1)}(z_0,z,z',J) \cr
&& + B(z_0,z)\,W_{n+1}^{(g)}(z',J)+ W_{n+1}^{(g)}(z,J)\,B(z_0,z') \cr
&& + \sum_{h',I'\subset J} W_{2+\#I'}^{(h')}(z_0,z,I')\,W_{1+n-\#I'}^{(g-h')}(z',J\setminus I') \cr
&& + \sum_{h',I'\subset J} W_{1+\#I'}^{(h')}(z,I')\,W_{2+n-\#I'}^{(g-h')}(z_0,z',J\setminus I') \Big]
\eea
which proves our recursion hypothesis to order $2g-2+n+1$.

\bibliography{biblio}
\bibliographystyle{plain}

\end{document}